\documentclass[prl,twocolumn,showpacs,superscriptaddress,floatfix]
{revtex4}
\usepackage{graphics,epsfig,amsfonts,amssymb,amsmath,ulem,color}

\newcommand{\beginsupplement}{%
        \setcounter{table}{0}
        \renewcommand{\thetable}{S\arabic{table}}%
        \setcounter{figure}{0}
        \renewcommand{\thefigure}{S\arabic{figure}}%
     }

\begin{document}
\title{Strong electronic correlations and Fermi surface reconstruction in the quasi-one dimensional iron superconductor BaFe$_2$S$_3$}
\author{J.M. Pizarro}
\author{E. Bascones}
\email{leni.bascones@icmm.csic.es}
\affiliation{Materials Science Factory. Instituto de Ciencia de Materiales de Madrid, 
ICMM-CSIC, Cantoblanco, E-28049 Madrid (Spain).}

\date{\today}
\begin{abstract}  
BaFe$_2$S$_3$ is a special iron superconductor with two-leg ladder structure which can help to unravel the 
role played by the electronic correlations in high-Tc superconductivity.  At zero pressure it 
is insulating with stripe antiferromagnetic (AF) order and superconductivity emerges under pressure. We use a slave-spin technique to analyze the strength of the local correlations in BaFe$_2$S$_3$. We find that at the pressure at which the superconductivity appears the electronic correlations in  BaFe$_2$S$_3$ are similar to 
the ones measured in other iron superconductors. At zero pressure the strength of the correlations is strongly enhanced being 
particularly severe for the two orbitals with the largest weight at the Fermi level what invalidates nesting as the mechanism for AF. At zero temperature the system is not a Mott insulator, but these two orbitals with mass enhancements $m^* \sim 12-15$ will become incoherent at higher temperatures. 
 Different from what happens in other iron superconductors, at both pressures,  
the Fermi surface is  reconstructed by the electronic correlations.  
\end{abstract} 
\maketitle
Ten years after the discovery of  high-Tc superconductivity in iron materials its origin is not 
understood\citep{kamihara08}. AF interactions are believed to play a key role in the emergence of 
superconductivity. Whether the magnetism is due to a Fermi surface instability\citep{mazin-PRL2008} of itinerant carriers, to exchange interactions between localized electrons
\citep{qimiaosi_natrev2016} or to double exchange physics\citep{nosotras_prb2012-2} is 
controversial\citep{nosotras_review2016}. 
\begin{figure}
%\leavevmode
%\hskip -1.0cm
\vspace{-0.9 cm}
\includegraphics[clip,width=0.33\textwidth]{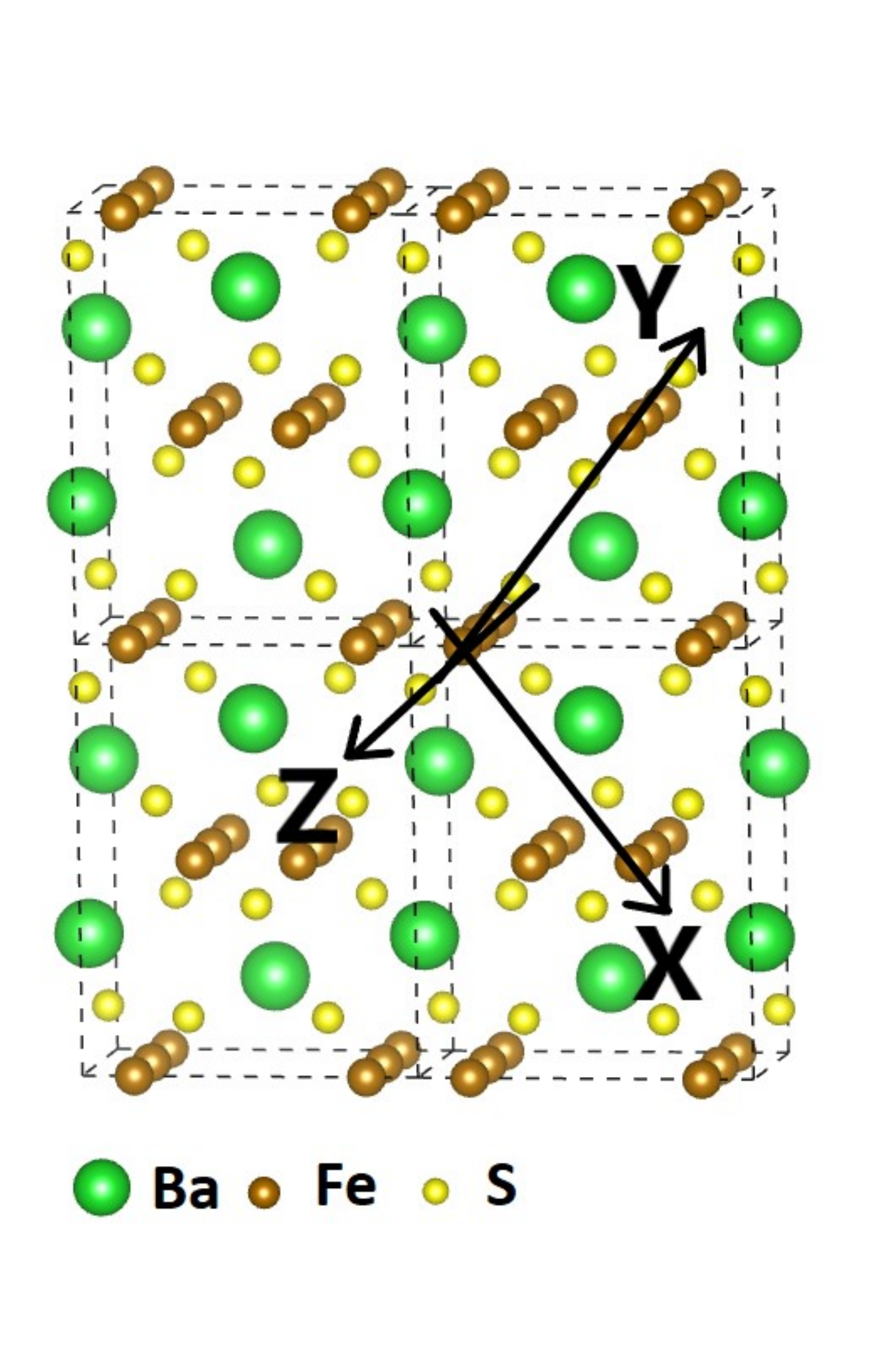}
\vspace{-1.0 cm}
\caption{(Color online) Crystal structure of BaFe$_2$S$_3$ with planes containing iron two-leg ladders. The 
iron-sulfur plane structure is equivalent to the one in iron superconductors except that one of 
each three 
iron rows is missing. The position of the missing row shifts between consecutive planes.} 
\label{fig:Figlattice} 
\vspace{-0.3 cm}
\end{figure}

Well studied iron superconductors have a planar square lattice. Recently superconductivity 
has been found in the two-leg ladder 123-compounds BaFe$_2$S$_3$ and BaFe$_2$Se$_3$  up to 24 K and 11 K respectively\citep{takahashi_natmat2015,yamauchi_prl2015,Struzhkin_prb2017}. 
As in planar iron superconductors, in these compounds Fe is tetrahedrally coordinated, but every third Fe row in the 
layer is missing, see Fig.\ref{fig:Figlattice}.  The nominal occupation of Fe, 6 
electrons in 5 orbitals, equals the one in  {\it undoped} iron superconductors.  However these 
ladder materials are AF insulators at zero pressure and superconductivity emerges under pressure. 
In BaFe$_2$Se$_3$ the ladders are tilted and the iron atomic distances distort as a plaquette AF 
ordering with  $\mu=2.8 \mu_B$ emerges
\citep{mcqueen_prb2011,medvedev_jetp2012,dagotto_prb2012}. Pressure suppresses the tilting of the 
ladders, induces a metal-insulator transition and probably a change to AF stripe order  before 
superconductivity appears\citep{svitlyk_jpcm2013,Struzhkin_prb2017,dagotto_prb2018}.  

BaFe$_2$S$_3$ shows stripe  order, AF along the ladder and 
ferromagnetic along the rung, with Neel temperature and magnetic 
moment $T_N \sim 120$ K and $\mu \sim 1-1.2 \mu_B$, close to the ones of other iron superconductors \citep{takahashi_natmat2015,yamauchi_prl2015,ohgushi_prl2016,birgueneau_prb2017}. Insulating character is observed above and below $T_N$. A resistivity anomaly around 
$180$ K could indicate orbital order\cite{yamauchi_prl2015,birgueneau_prb2017}.  Metallicity and superconductivity appear 
at $P\sim 11$ GPa. Ab-initio calculations reproduce the AF order but with larger moment $\mu \sim 2.1 \mu_B$\citep{arita_prb2015-abinitio,dagotto_prb2017} and nesting arguments have been used to explain intraladder AF\citep{arita_prb2015-downfolding}. Other authors have suggested that BaFe$_2$S$_3$ is 
is a Mott insulator in which pressure suppress the Mott gap and induce 
superconductivity\citep{takahashi_natmat2015,yamauchi_prl2015}. However, the activation gap is only $70 $ meV and the magnetic moment is much smaller than expected from saturated spins 
$\mu=4.0 \mu_B$\citep{ohgushi_prl2016}. Photoemission and X-ray experiments suggest the presence 
of local and itinerant electrons \citep{ootsuki_prb2015,takubo_prb2017}.

In this paper we use a slave spin technique to study the strength of electronic correlations in 
BaFe$_2$S$_3$ and shed light on the nature of the  superconducting and AF instabilities. 
We find that at pressures for which superconductivity appears the electronic correlations 
are similar to those found in other iron superconductors. However at zero pressure the 
correlations are much stronger, especially for the orbitals with the largest weight at the Fermi level. 
Such strong correlations invalidate nesting as the origin of AF in this material with local and itinerant electrons. We also show that electronic correlations 
reconstruct the Fermi surface at both pressures.     

{\it Model and methods} We consider a multi-orbital model for Fe atoms. Interactions, restricted to orbitals in the same 
atom, include intraorbital $U$, interorbital $U'$, Hund's coupling $J_H$, and pair hopping $J'$ 
terms, see the supplementary information (SI) and \cite{nosotras_review2016}. We start from the tight-binding models for BaFe$_2$S$_3$ derived from ab-initio 
calculations at pressures $P=0$ and $P=12.4$ GPa\cite{arita_prb2015-downfolding} . They include 20 orbitals, 5 per 
each of the four iron atoms in the unit cell. Orbital and unit cell axis differ among them and from the ones commonly used in iron superconductors. X and Y axis connect ladders in adjacent planes and Z runs along 
the ladders, see Fig.\ref{fig:Figlattice}. The orbitals $zx$, $yz$, $xy$ and $3z^2-r^2$, $x^2-y^2$ 
are defined with $z$ along the ladders, $x$ 
connecting ladders in the same Fe plane and $y$ axis perpendicular to the Fe-ladders plane.  
In this basis there are finite onsite non-diagonal terms among these orbitals. We change to a basis with on-site diagonal terms only
w$_\alpha$=w$_{zx}$, w$_{yz}$, w$_{xy}$, w$_{3z^2-r^2}$, w$_{x^2-y^2}$. Here the subscript $\alpha$ indicates which orbital in the original basis has the largest weight, see SI. 

The interaction terms are defined in the w$_\gamma$ basis. We take atomic filling $n=6$, $U'=U-2J_H$, $J'=J_H$, assume 
$J_H=0.25 U$ and study the electronic correlations as a function of $U$\cite{castellani78}. We quantify the strength of 
the  local correlations  by the orbital dependent quasiparticle weight $Z_\gamma$ calculated with the U(1) slave-spin technique\cite{yu_prb2012}. $Z_\gamma=1$  in non-correlated materials and decreases 
as the electronic correlations increase. It has the same value in the four Fe atoms of the unit cell. In 
the approximation used, $Z_\gamma$ equals the inverse of the mass enhancement factor of each orbital $m^*_\gamma$. 

Using 
constraint RPA the strength of the interactions in BaFe$_2$S$_3$ was estimated to be similar to that in 
LiFeAs, i.e. larger than the interaction in BaFe$_2$As$_2$ and smaller than in FeSe, and to be reduced a 6$\%$ 
under a pressure of $12.4$ GPa\cite{arita_prb2015-downfolding, miyake2010}. Slave-spin calculations  for BaFe$_2$As$_2$ and FeSe using $J_H=0.25 U$ 
compare well with experiment if the interactions $U_{BaFe_2As_2}=2.7$ eV and $U_{FeSe}=3.0$ eV are used\cite{demedici_prl2014,fanfarillo_prb2017}. 
Therefore to study the electronic correlations in BaFe$_2$S$_3$ we take $U_{P=0}=2.90$ eV for $P=0$ and $U_{P=12.4}=2.75$ eV for $P=12.4$ GPa.

\begin{figure}
\leavevmode
%\hskip -1.0cm
\includegraphics[clip,width=0.42\textwidth]{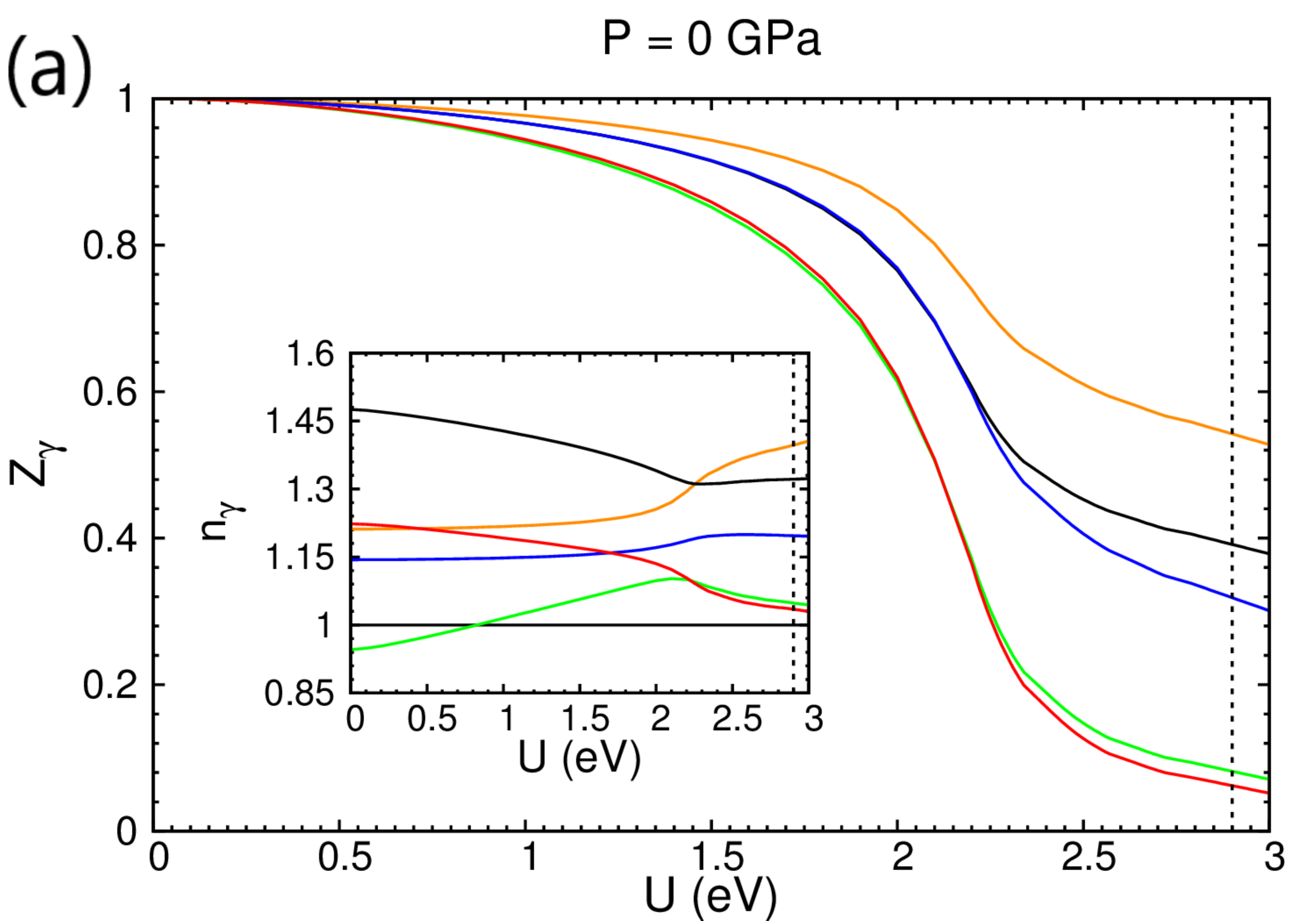}
\includegraphics[clip,width=0.42\textwidth]{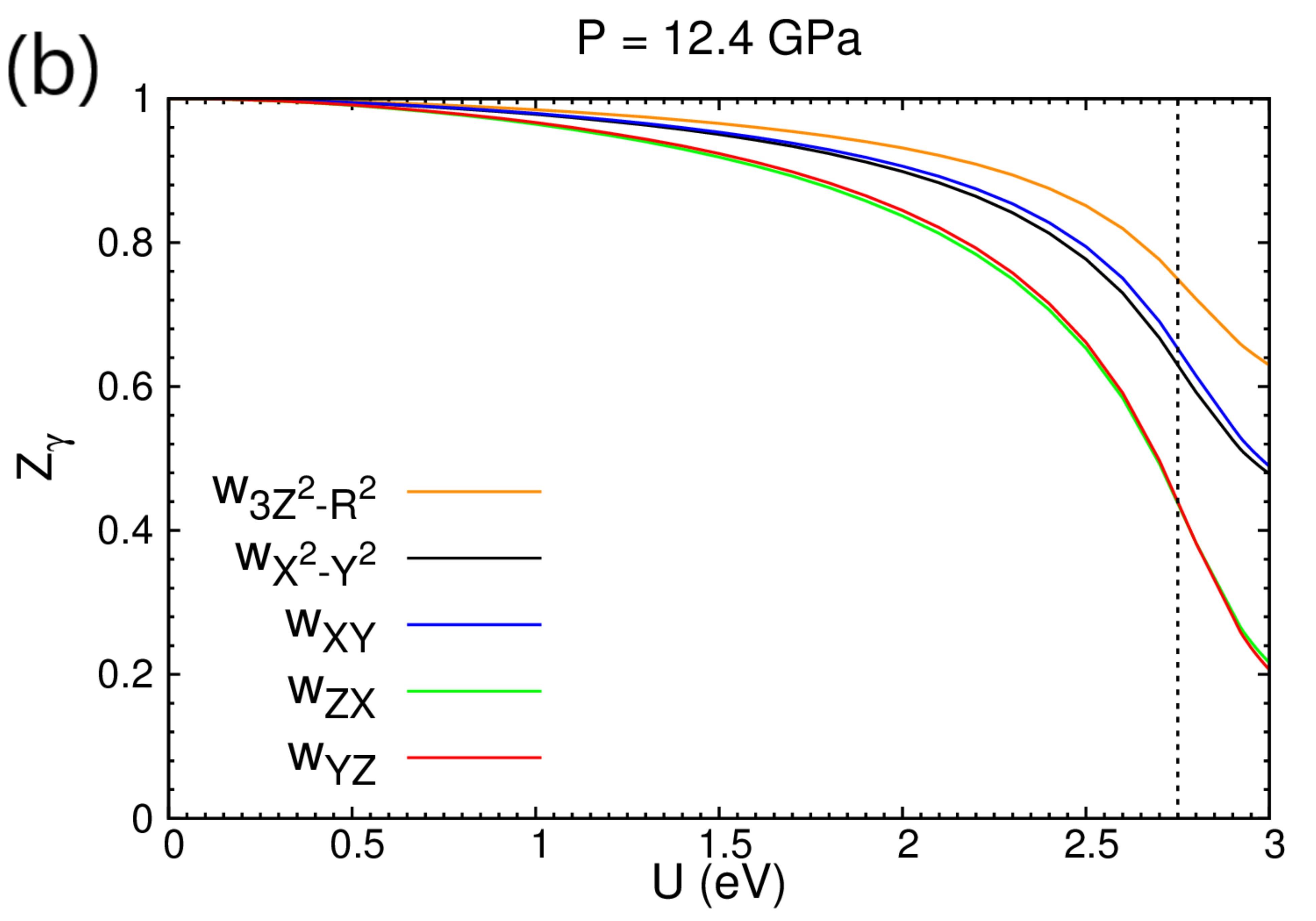}
\caption{(Color online) (a)Orbital dependent quasiparticle weight $Z_\gamma$ (main figure) and filling  $n_\gamma$ 
(inset) at zero pressure versus the intraorbital interaction $U$.  At the Hund's metal crossover, 
$U^* \sim 2.1$ eV,  the system becomes strongly correlated due to the formation of local spins and a sharp drop of $Z_\gamma$ is 
observed . The dotted line marks the interaction value $U_{P=0}=2.9$ eV suitable 
for BaFe$_2$S$_3$, see text. At this interaction the  orbitals $w_{yz}$ and $w_{zx}$  with have the largest weight the bands close to the Fermi 
level are near half-filling and have a very small quasiparticle weight $Z_\gamma$  which corresponds to 
an enhanced mass $m^* \sim 15$.  (b) Same as in main figure in (a) for P=12.4 GPa, pressure at which 
superconductivity is found. With pressure the Hund's metal crossover shifts to larger interactions due to 
larger bandwidth and the interaction is reduced to $U_{P=12.4}\sim 2.75$ eV.  At this interaction the 
strength of the correlations, as measured by $Z_\gamma$,  becomes similar to the one found in other iron 
superconductors. }  
\label{fig:FigZ} 
\vspace{-0.3cm}
\end{figure}
\begin{figure*}
\leavevmode
%\hskip -1.0cm
\includegraphics[clip,width=0.42\textwidth]{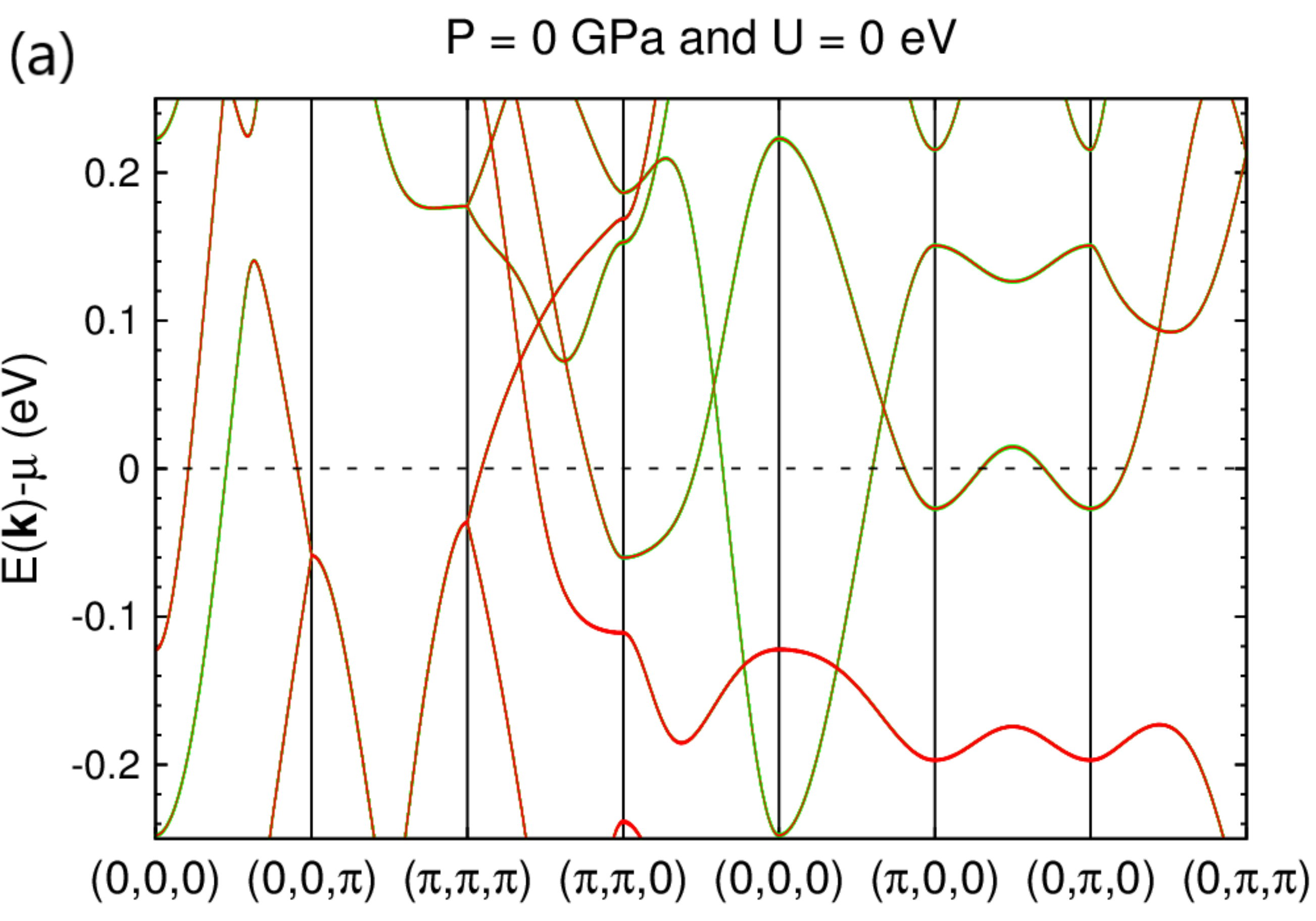}
\includegraphics[clip,width=0.42\textwidth]{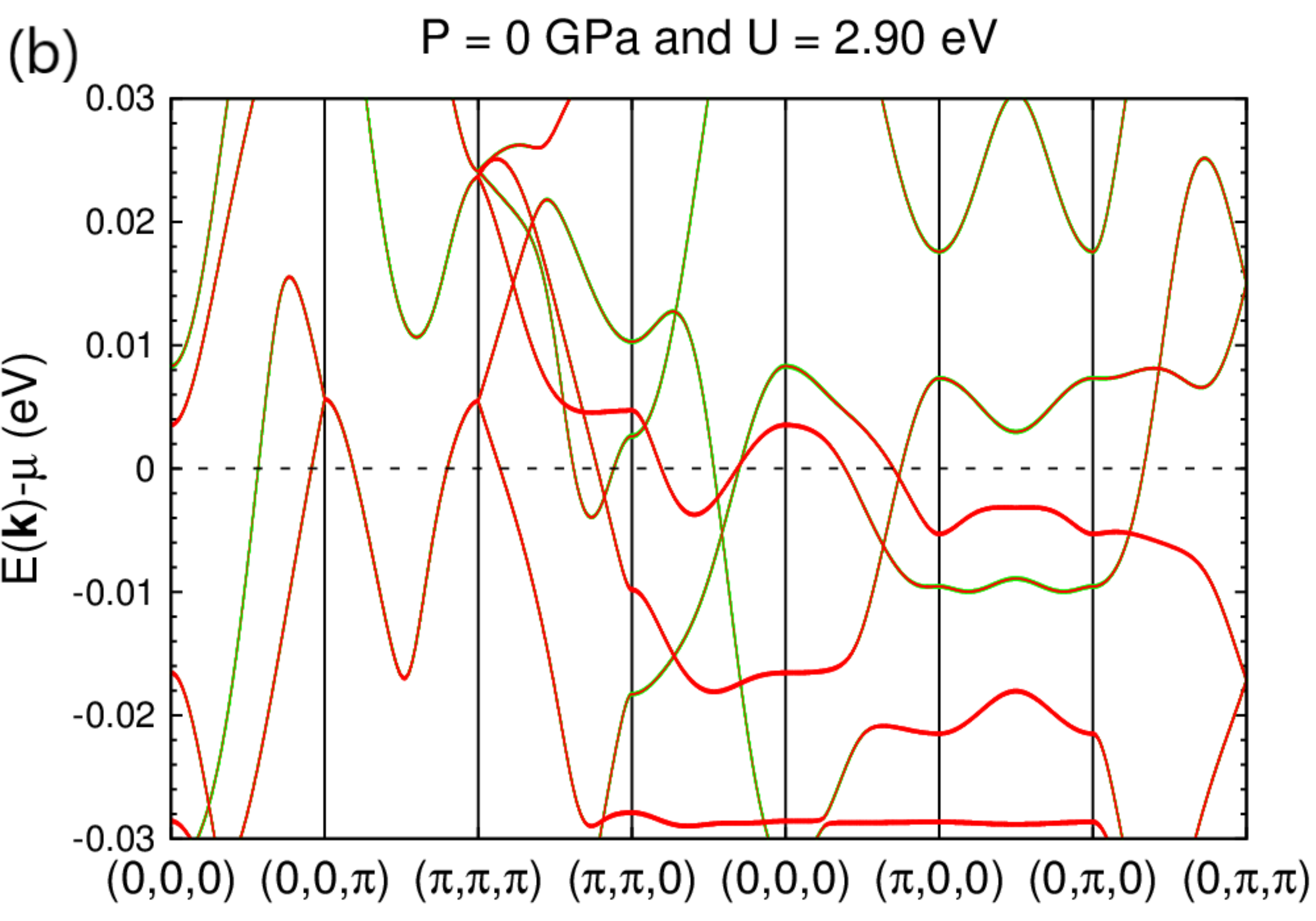}
\includegraphics[clip,width=0.42\textwidth]{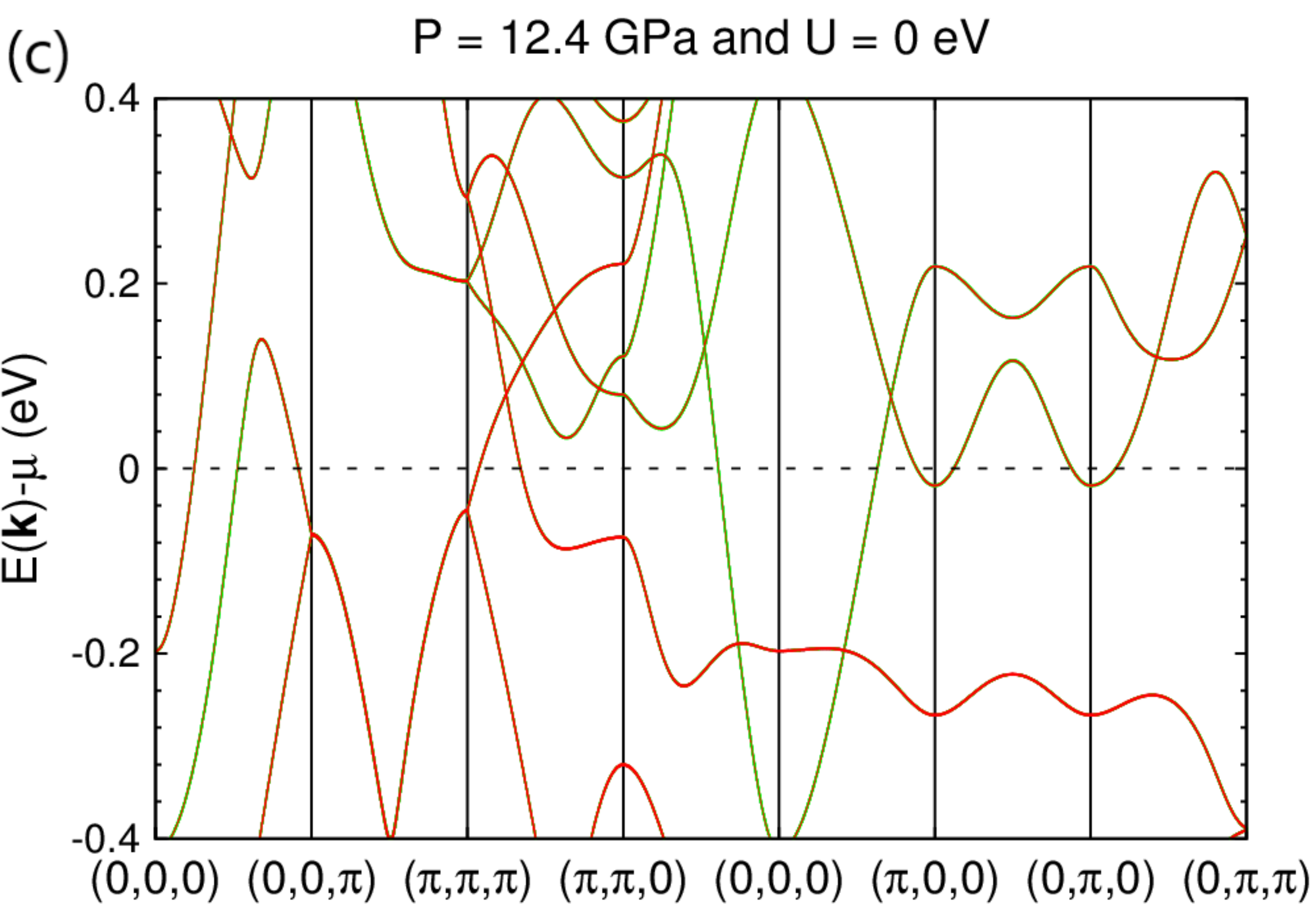}
\includegraphics[clip,width=0.42\textwidth]{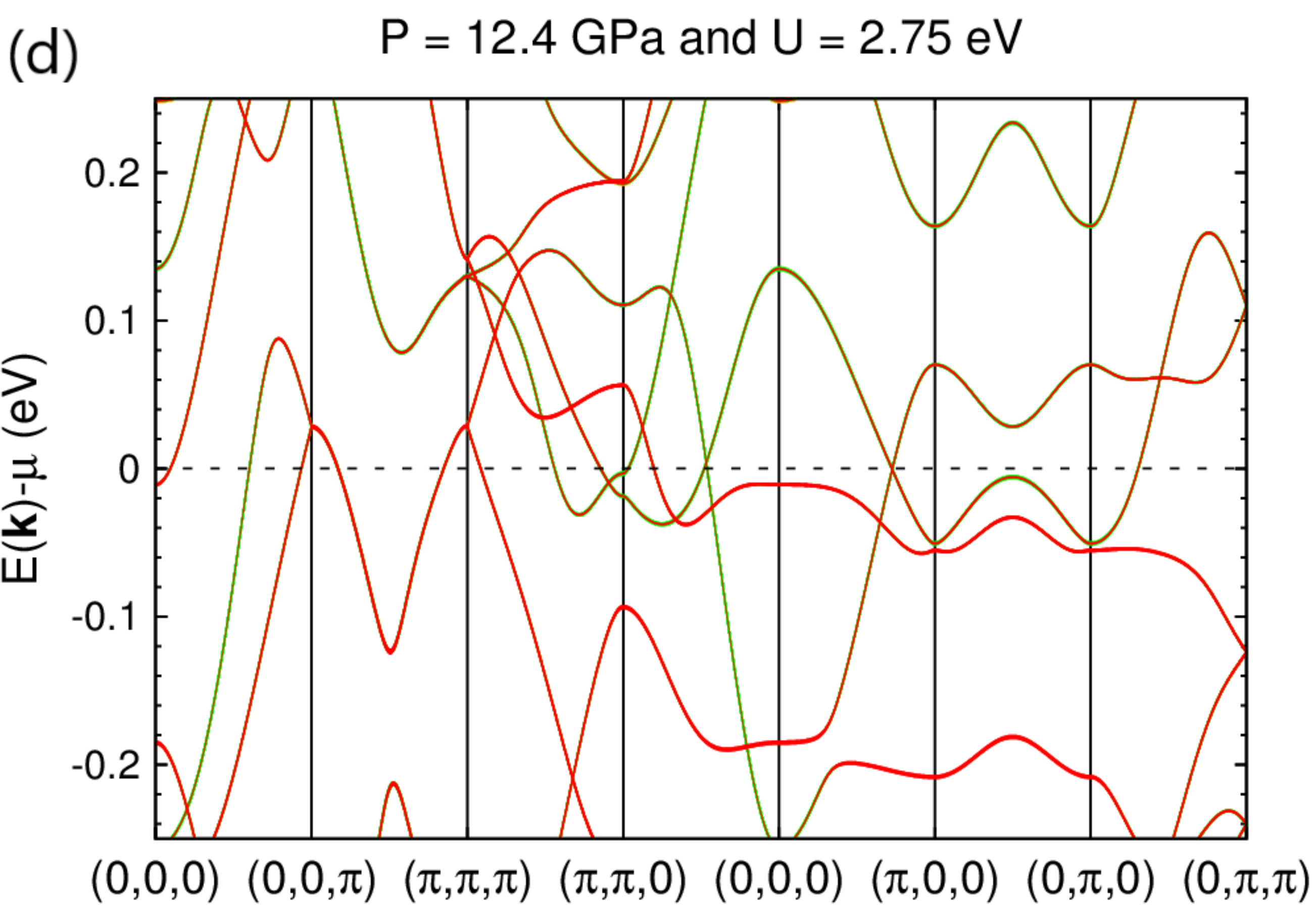}
\caption{(Color online) Band structure near the Fermi level for $P=0$ and $P=12.4$ 
GPa with and without including the electronic correlations. Close to the Fermi level the bands are 
dominated by w$_{yz}$ (red) and w$_{zx}$ (green) orbitals. The electronic correlations modify the band structure and reconstruct the Fermi surface, characterized by very shallow pockets.   } 
\label{fig:Figzoombandas} 
\vspace{-0.3cm}
\end{figure*}
 {\it Electronic correlations} Fig.\ref{fig:FigZ} shows the  quasiparticle weight $Z_\gamma$ as a function of $U$.  At $P=0$, $Z_\gamma$  is strongly suppressed at $U^*\sim 2.1$ eV. Beyond this crossover the system enters into a correlated state, the Hund's metal, with  well formed local spins satisfying Hund's rule\cite{nosotras_review2016, werner2008, haule09, demedici_prb2011, demedici11, yu_prb2012,fanfarillo_prb2015}.  Two of the orbitals, w$_{zx}$ and w$_{yz}$, become very  strongly correlated, $Z_\gamma < 0.1$, while the other three orbitals 
w$_{xy}$, w$_{3z^2-r^2}$ and w$_{x^2-y^2}$ show intermediate correlations $Z_\gamma \sim 0.3-0.6$. The suppression of the quasiparticle weight $Z_\gamma$ is concomitant with a 
reorganization of the orbital filling $n_\gamma$, see 
inset in Fig.\ref{fig:FigZ} (a). The strongly correlated orbitals w$_{zx}$ and 
w$_{yz}$ approach half-filling in the Hund's metal state. The interaction at which the Hund's metal crossover appears in Fig.\ref{fig:FigZ}(a) is reduced 
with respect to the one found in other iron superconductors $U^*\sim 2.6-2.7$ eV \citep{yu_prb2012,demedici_prl2014,fanfarillo_prb2017}. Behind this reduction it is the smaller 
value of the bandwidth: 4 eV in BaFe$_2$S$_3$, see\citep{arita_prb2015-downfolding} and SI,  and close to 5 eV in other compounds. 

BaFe$_2$S$_3$ with $U_{P=0}=2.90$ eV  is a strongly correlated Hund metal. At this interaction, the electronic correlations reduce the bandwidth to less than 2 eV, see SI.  The narrowing of the bands is especially prominent close to the Fermi level. This happens because the bands close to the Fermi level are 
dominated by w$_{zx}$ and w$_{yz}$ orbitals with a very small $Z_\gamma$ which results in mass enhancement factors $\sim 12-15$, Fig.\ref{fig:Figzoombandas}(a) and (b).

At $P=12.4$ GPa the strength of the electronic correlations is  reduced with respect to its zero pressure values. The quasiparticle weight and mass enhancement factors of the orbitals close to the Fermi level  $m^*\sim 2-3$ are similar to 
the ones found in the planar iron superconductors\citep{nosotras_review2016, mingyi_npjquantum2017}. This result is especially interesting because BaFe$_2$S$_3$ is superconducting at $P=12.4$ GPa . The Hund's metal crossover  
is shifted from $U\sim 2.1$ eV at $P=0$ GPa to  $U\sim 2.7$ eV at $P=12.4$ GPa eV, due to the larger bandwidth $\sim 5$ eV see  SI, what places BaFe$_2$S$_3$
at the Hund's metal crossover at $P=12.4$ GPa.

{\it Fermi surface reconstruction} Besides the flattening of the bands, the Fermi surface is reconstructed at both pressures. This is different from what happens in other iron superconductors for which the inclusion of local correlations does not alter significatively the Fermi surface shape.  Close to the Fermi level the 
reduced hybridization with other orbitals shifts w$_{yz}$ upwards and w$_{zx}$ downwards producing the Fermi surface reconstruction. 

\begin{figure*}
\leavevmode
%\hskip -1.0cm
\includegraphics[clip,width=0.43\textwidth]{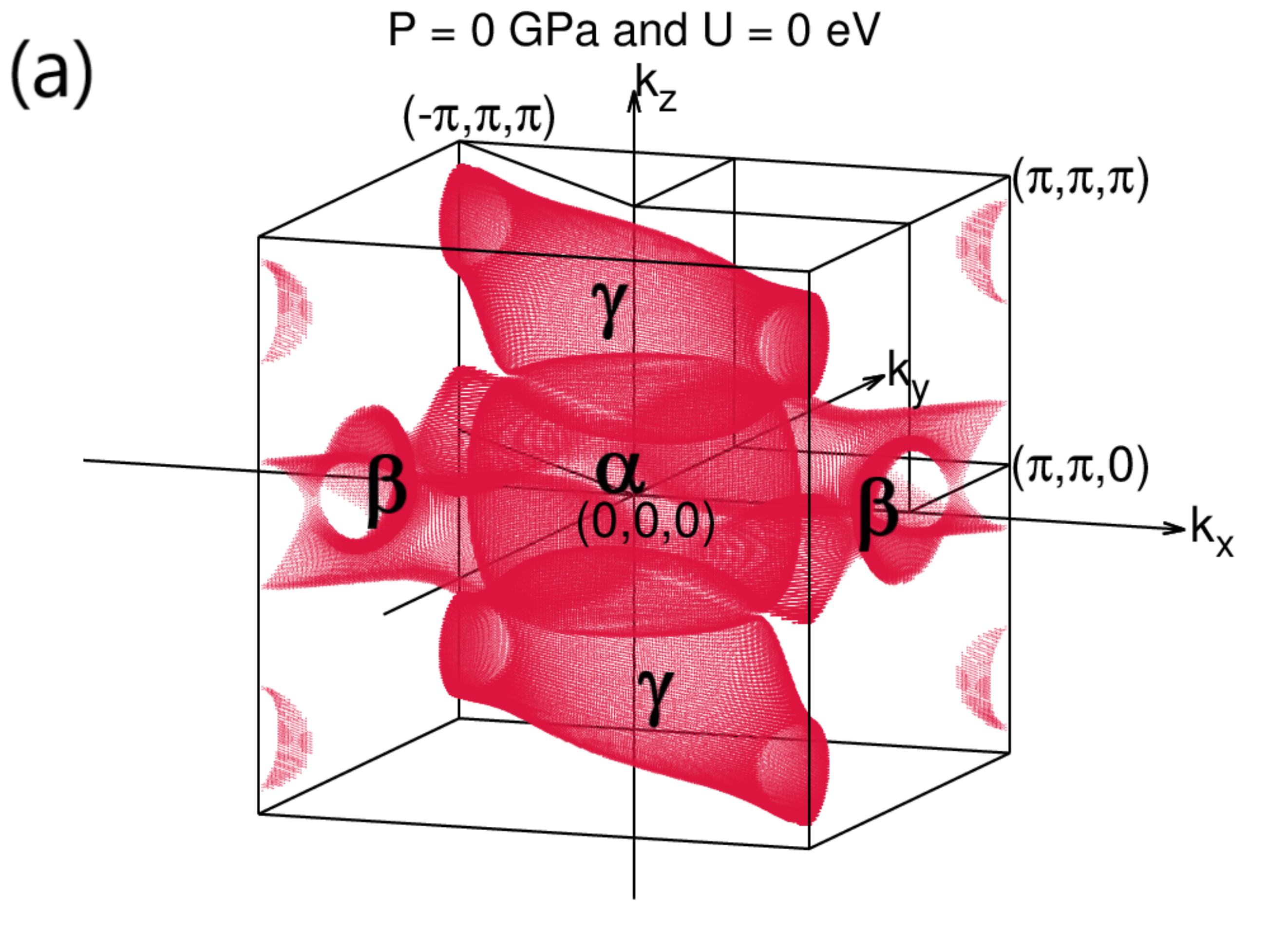}
\includegraphics[clip,width=0.43\textwidth]{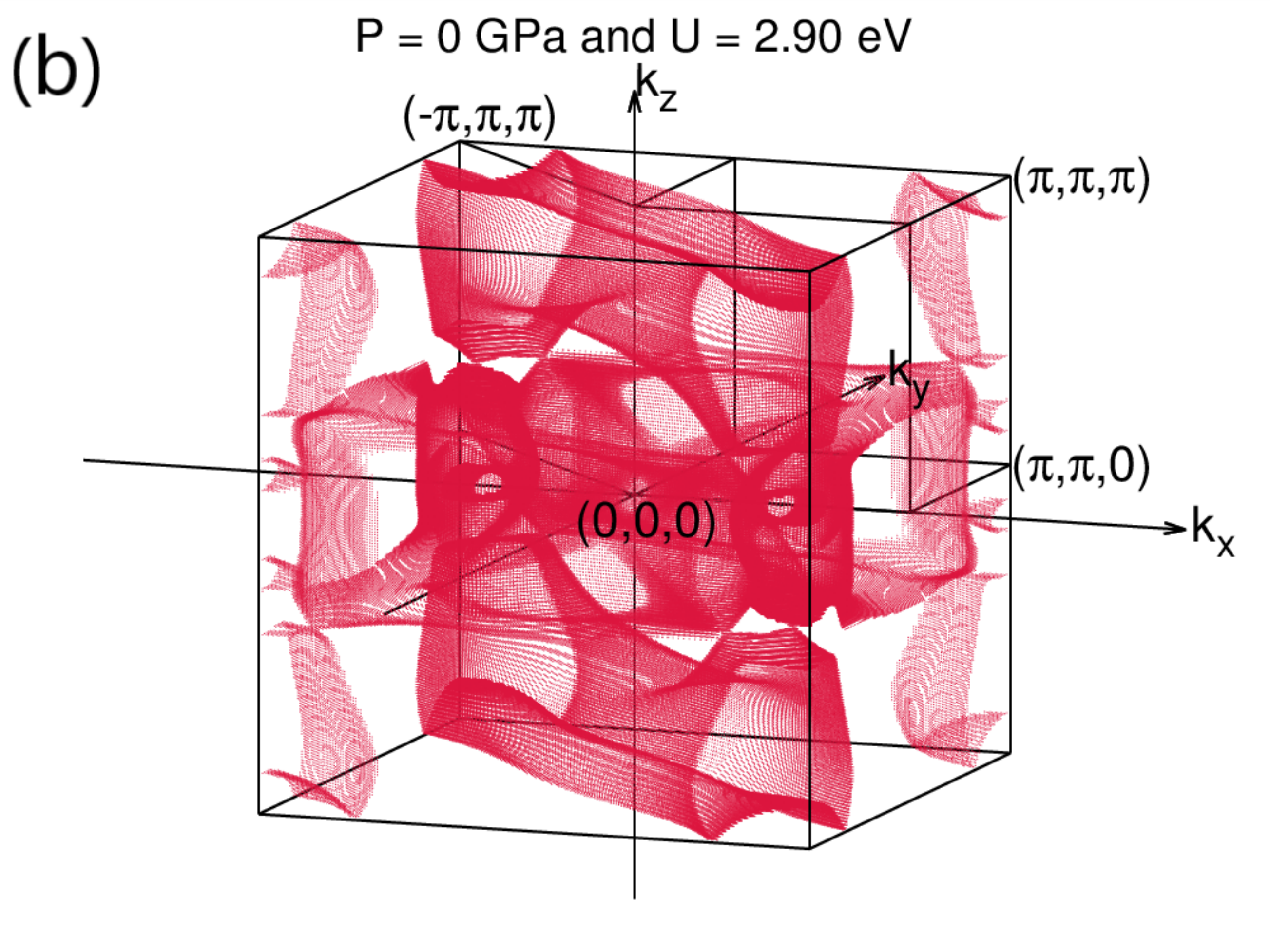}
\includegraphics[clip,width=0.43\textwidth]{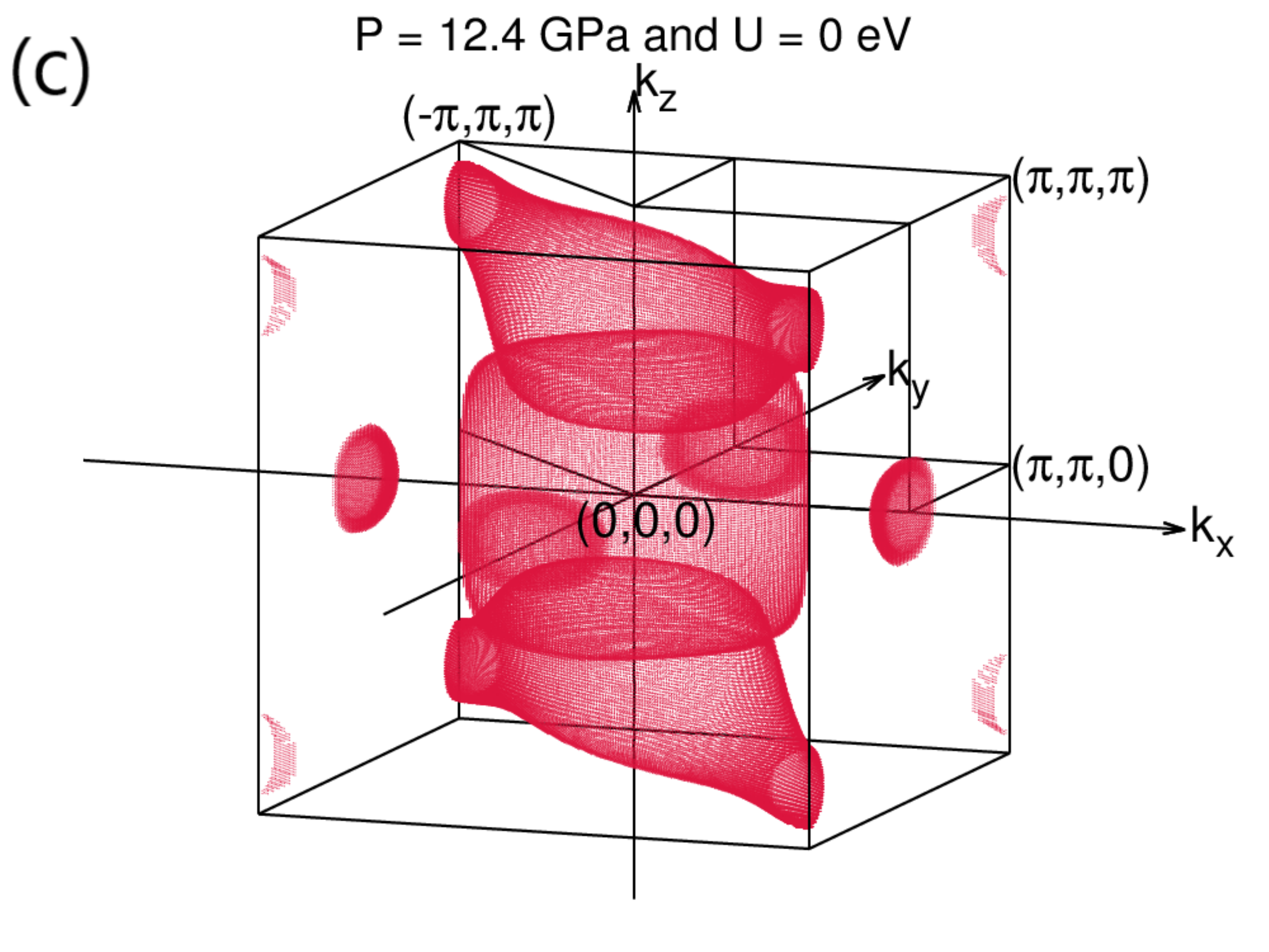}
\includegraphics[clip,width=0.43\textwidth]{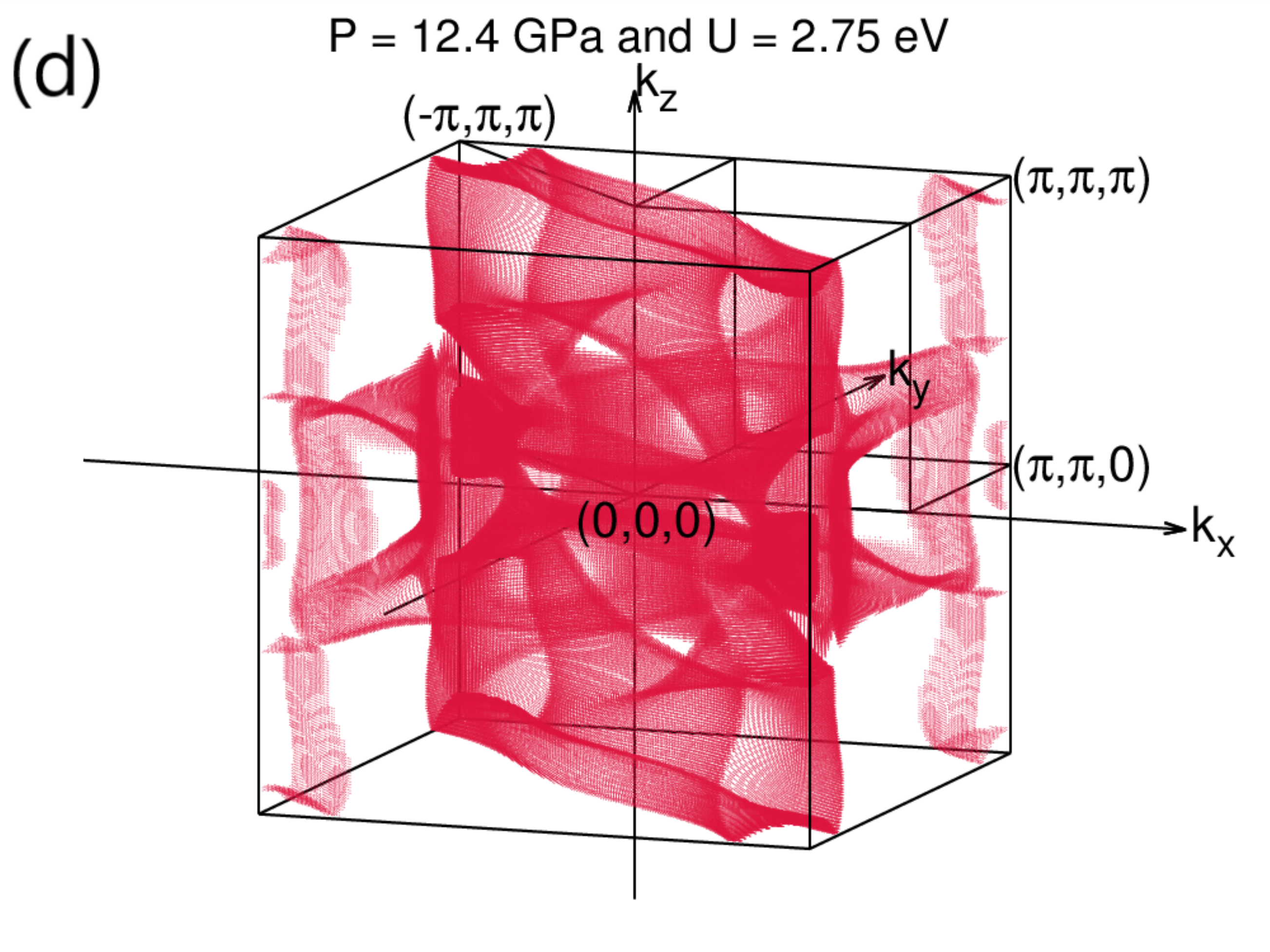}
\caption{(Color online)Reconstruction of the Fermi surface due to the electronic correlations at 
$P=0$ GPa and $P=12.4$ GPa. On spite of the quasi-one dimensional lattice the Fermi surface has three dimensional character. When correlations are included enhanced scattering at $\bf Q \sim (0,0,2\pi)$ ($\bf Q^* \sim (0,0,\pi)$ if the Brillouin zone is unfolded along $k_z$) between
hole pockets along $(k_x,-k_x,\pm \pi)$  is expected.} 
\label{fig:Figfs} 
\vspace{-0.3cm}
\end{figure*}

In the absence of correlations the ab-initio Fermi surface has electron 
pockets $\alpha$ and $\beta$ respectively centered at $\Gamma$ and along $(\pm \pi,k_y,0)$ and $(k_x,\pm \pi,0)$  
and hole pockets $\gamma$ around $(0,0,3/4\pi)$ 
directed along the $(k_x, -k_x, \pm 3/4 \pi)$, see Fig.\ref{fig:Figfs}, SI and Ref\citep{arita_prb2015-downfolding}.  With pressure the size of the pockets change, especially the $\beta$ pockets which shrinks significatively at $P=12.4$ GPa.

When the effect of correlations is included the $\gamma$ hole pockets change their shape and cut $k_z=\pm \pi$, the $\beta$ pockets  grow and new smaller hole  pockets appear close to $(\pi,\pi,0)$. At $P=0$ the $\alpha$ pocket disappears  at $\Gamma$  but 
smaller triangular-like electron pockets still remain around $(\pi/2,\pi/2,0)$.   At $P=12.4$ GPa the $\alpha$ pocket   is still present  at $\Gamma$ but with an "H" form, see SI . The large mass enhacements make all the Fermi pockets extremely shallow at $P=0$. The top and bottom of the hole and electron bands which give rise to the 
Fermi pockets are shifted only a few meV of the Fermi level, see Fig. \ref{fig:Figzoombandas}.

{\it Summary and discussion} We find that at zero pressure BaFe$_2$S$_3$ is a strongly correlated Hund metal with orbital selective correlations. The two orbitals with the largest weight at the Fermi level have very small quasiparticle weight $Z_{w_{zx},w_{yz}}\sim 0.06$, i.e. large mass enhancements $m^*_{w_{zx},w_{yz}}\sim 12-15$ are expected. In the other three orbitals the strength of  
correlations and the mass enhancements are intermediate
$Z_m\sim 0.3-0.5$ and $m^*_m\sim 2-3.5$.  

At zero temperature BaFe$_2$S$_3$ is not a Mott insulator. Several bands cross the Fermi level and reconstruct the Fermi surface. However, due to the small quasiparticle 
weight, if the temperature is not very low the electrons close the Fermi level will behave as incoherent particles: the quasiparticle peak is expected to be absent in photoemission 
experiments and the resistivity could show insulating behavior as observed 
experimentally. Short range AF correlations above the Neel temperature will 
enhance the insulating tendencies localizing these very weakly coherent electrons. In agreement with photoemission and X-ray experiments the system can be described in terms of itinerant and localized electrons \citep{ootsuki_prb2015,takubo_prb2017}. 

The reconstructed Fermi surface shows enhanced scattering at $\bf Q \sim (0,0,2\pi)$ between the hole pockets. In a  Brillouin zone unfolded along $k_z$, $\bf Q$ would become $\bf Q^* \sim (0,0,\pi)$ coincident with the intra-ladder AF momentum. These pockets are very shallow and therefore very sensitive to other effects not included in this work. But overall the small quasiparticle weight prevents any explanation of the antiferromagnetic state in terms of nesting.  This can have important consequences for the understanding of the superconducting phase which it is believed to be mediated by the AF correlations. Whether the AF order in this material is better explained by a double exchange mechanism or with local moments is an interesting issue beyond the scope of this work. 

Interestingly, at $P=12.4$ GPa, at which the superconductivity is found, the strength of the correlations is reduced to values similar to the ones measured in other iron superconductors. Intermediate correlations could be beneficial for high-Tc temperature superconductivity in iron superconductors\citep{pizarro_prb2017}. 

As in the $P=0$ case, at $P=12.4$ GPa the reconstructed Fermi surface shows enhanced scattering at $\bf Q \sim (0,0,2\pi)$, $\bf Q^* \sim (0,0,\pi)$ in the unfolded Brillouin zone. At this pressures the Fermi pockets are not so shallow and the larger quasiparticle weight could, in principle, justify an explanation of the superconductivity in terms of the processes at the Fermi surface, or at least some contribution to it.  Such mechanism would predict an order parameter with sign change between the two hole-pockets. An interesting issue which requires further work is whether this weak coupling mechanism for superconductivity is invalidated by the non-nesting character of the AF phase suppressed by pressure or not.

In conclusion our work shows that the quasi-one dimensional 123 family can be very useful to decipher the nature of high-Tc superconductivity in iron based materials. Further experimental and theoretical work is desired.   

We thank E. Dagotto por useful conversations and exchange of data and M.J. Calder\'on and L. Fanfarillo for useful conversations. Funding from Ministerio de Econom\'ia y Competitividad via grant No. FIS2014-53218-P, and from Fundaci\'on Ram\'on Areces is gratefully acknowledged. 
\bibliography{bafe2s3}
\newpage
\beginsupplement
\section{SUPPLEMENTARY INFORMATION}
\subsection*{Methods}
We start from a multi-orbital model with local interactions including: intraorbital $U$, interorbital $U'$, 
Hund's coupling $J_H$, and pair hopping $J'$ terms,

\begin{eqnarray}
\nonumber
  H   = \sum_{k,\gamma,\beta,\sigma}\epsilon_{k,\gamma,\beta}c^\dagger_{k,\gamma,\sigma}c_{k,\beta,\sigma}+h.c. + \sum_{j,\gamma,\sigma}\epsilon_\gamma n_{j,\gamma,\sigma}
\\ \nonumber
 +  U\sum_{j,\gamma}n_{j,\gamma,\uparrow}n_{j,\gamma,\downarrow}
 +  (U'-\frac{J_H}{2})\sum_{j,\{\gamma>\beta\},\sigma,\tilde{\sigma}}n_{j,\gamma,\sigma}n_{j,\beta,\tilde{\sigma}}
\\ 
 -  2J_H\sum_{j,\{\gamma >\beta\}}\vec{S}_{j,\gamma}\vec{S}_{j,\beta}
 +   J'\sum_{j,\{\gamma\neq
  \beta\}}c^\dagger_{j,\gamma,\uparrow}c^\dagger_{j,\gamma,\downarrow}c_{j,\beta,\downarrow}c_{j,\beta,\uparrow}
 \,
\label{eq:hamiltoniano}
\end{eqnarray}
$j$ label the Fe atoms. Each unit cell contains 4 atoms.  X and Y axis connect ladders in adjacent planes and Z runs along 
the ladders. $k$ is the momentum in the 4 Fe Brillouin zone, $\sigma$ 
the spin and $\gamma$, and $\beta$ the orbitals. We include five orbitals per Fe atom. There are 20 orbitals in the unit-cell. Curly brackets $\{ \}$ in the sum subscript indicate that the sum is  restricted to orbitals in the same atom, i.e., both orbitals are between 1 and 5 or between 6 and 10, and so on. The model is defined in w$_\alpha$ orbital basis. In this basis there are no hybridization terms between different orbitals within the same atom, only a diagonal on-site contribution $\epsilon_\gamma$. We assume
$U'=U-2J_H$~\cite{castellani78} and $J'=J_H$, as in rotationally invariant systems, leaving only two independent interaction parameters, $U$ and $J_H$. We take $J_H=0.25 U$. 

To obtain $\epsilon_{k,\gamma,\beta}$ and $\epsilon_\gamma$ we start from the 20-orbital tight-binding models calculated in\citep{arita_prb2015-downfolding} from a Wannier projection of ab-initio results for BaFe$_2$Se$_3$ at $P=0$ and $P=12.4$ GPa. These models are obtained using as orbital basis: $zx$, $yz$, $x^2-y^2$, $3z^2-r^2$ and $xy$ defined with $z$ along the ladders, $x$ connecting ladders in the same Fe plane and $y$ axis perpendicular to Fe-ladders plane. $zx$ and $xy$ along these axis are respectively equivalent to $xy$ and $zx$ in the basis frequently used in tight-binding models for iron superconductors (with x and y along Fe bonds and z perpendicular to the Fe plane), i.e. the two orbitals are exchanged; $yz$ is equivalent in both basis and $x^2-y^2$ and $3z^2-r^2$ do not have direct analogues. They are linear combinations of the orbitals defined with the axis exchanged.

In this basis there are inter-orbital onsite terms. We perform a change to a basis w$_{\alpha}$ in which the onsite terms of these tight-binding models are diagonal.  The subscript $\alpha$ corresponds to the orbital of the original basis which gives a larger contribution to w$_{\alpha}$ The matrix elements of the change of basis have the same absolute value in the four atoms, but their sign can differ. Their value depend on pressure. An example is given below. For $P=0$:

\begin{eqnarray}
\nonumber
w^1_{3z^2-r^2}= 0.95|3z^2-r^2 >  + 0.16|yz>   - 0.25|x^2-y^2>
\\ \nonumber
w^3_{3z^2-r^2} =0.95|3z^2-r^2 > -0.16|yz>   -0.25 |x^2-y^2>   
\end{eqnarray}
where the superscript label the atom in the unit cell. For $P=12.4$ GPa
\begin{eqnarray}
\nonumber
w^1_{3z^2-r^2} = 0.92|3z^2-r^2 >  + 0.22|yz>   - 0.33|x^2-y^2>
\\ \nonumber
w^3_{3z^2-r^2} = 0.92|3z^2-r^2 >  - 0.22|yz>   - 0.33|x^2-y^2>
\end{eqnarray}
We then write the onsite and hopping terms of the tight-binding models are the basis w$_\alpha$.To address the role of the electronic correlations we use $U(1)$ slave spin theory\citep{yu_prb2012} and keep only density-density terms. That is, pair hopping and spin-flip terms do not enter into the calculation. With this technique we calculate the quasiparticle weight and the onsite energy shifts which are generated by the electronic correlations.

\subsection*{Bands and Fermi surface reconstruction}
Below we provide some figures, complementary to Fig.\ref{fig:Figzoombandas} and  Fig.\ref{fig:Figfs} which show the effect of the electronic correlations in the band structure and in the Fermi surface at $P=0$ and $P=12.4$ GPa.
\begin{figure*}
\leavevmode
%\hskip -1.0cm
\includegraphics[clip,width=0.42\textwidth]{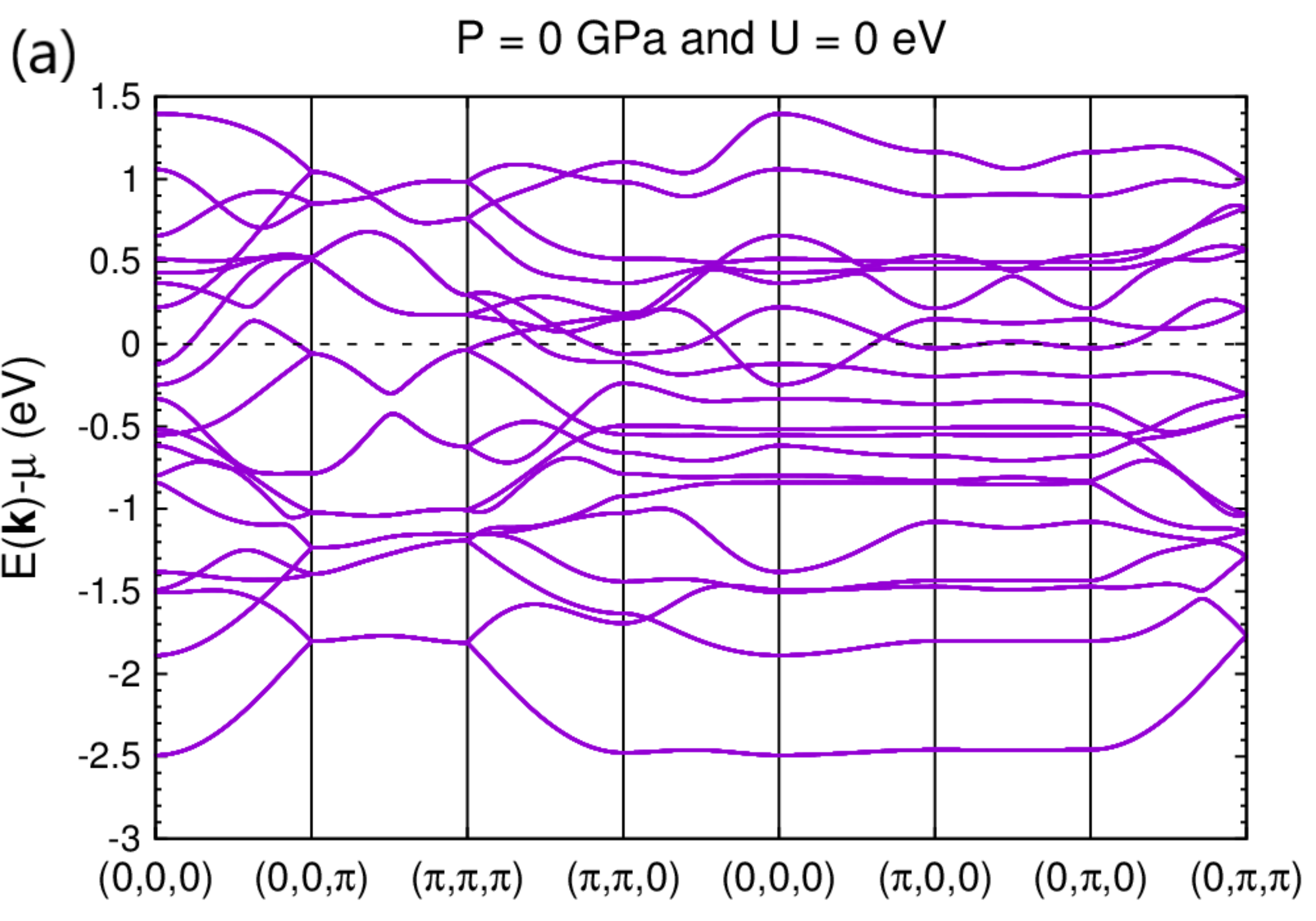} 
\includegraphics[clip,width=0.42\textwidth]{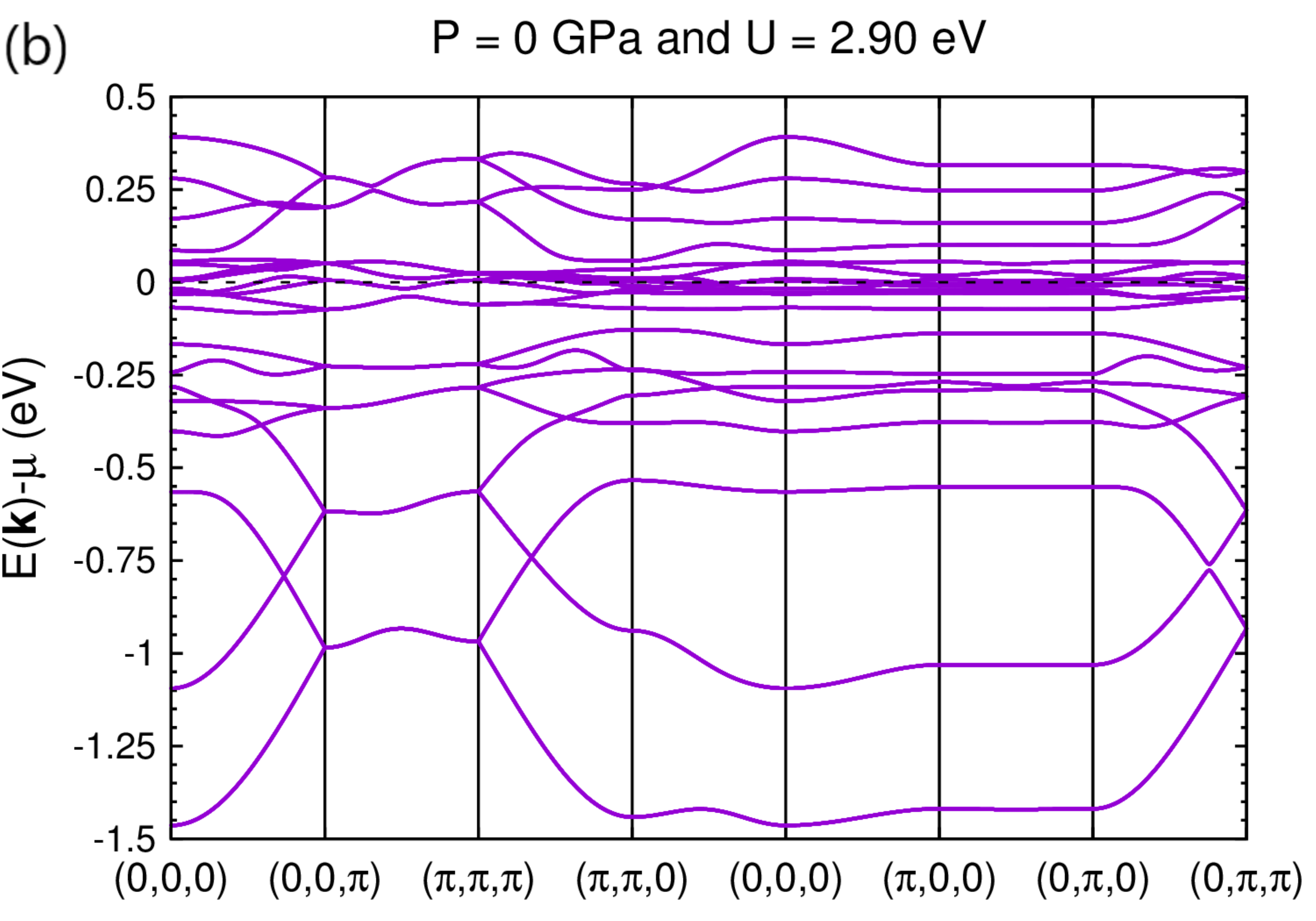} 
\includegraphics[clip,width=0.42\textwidth]{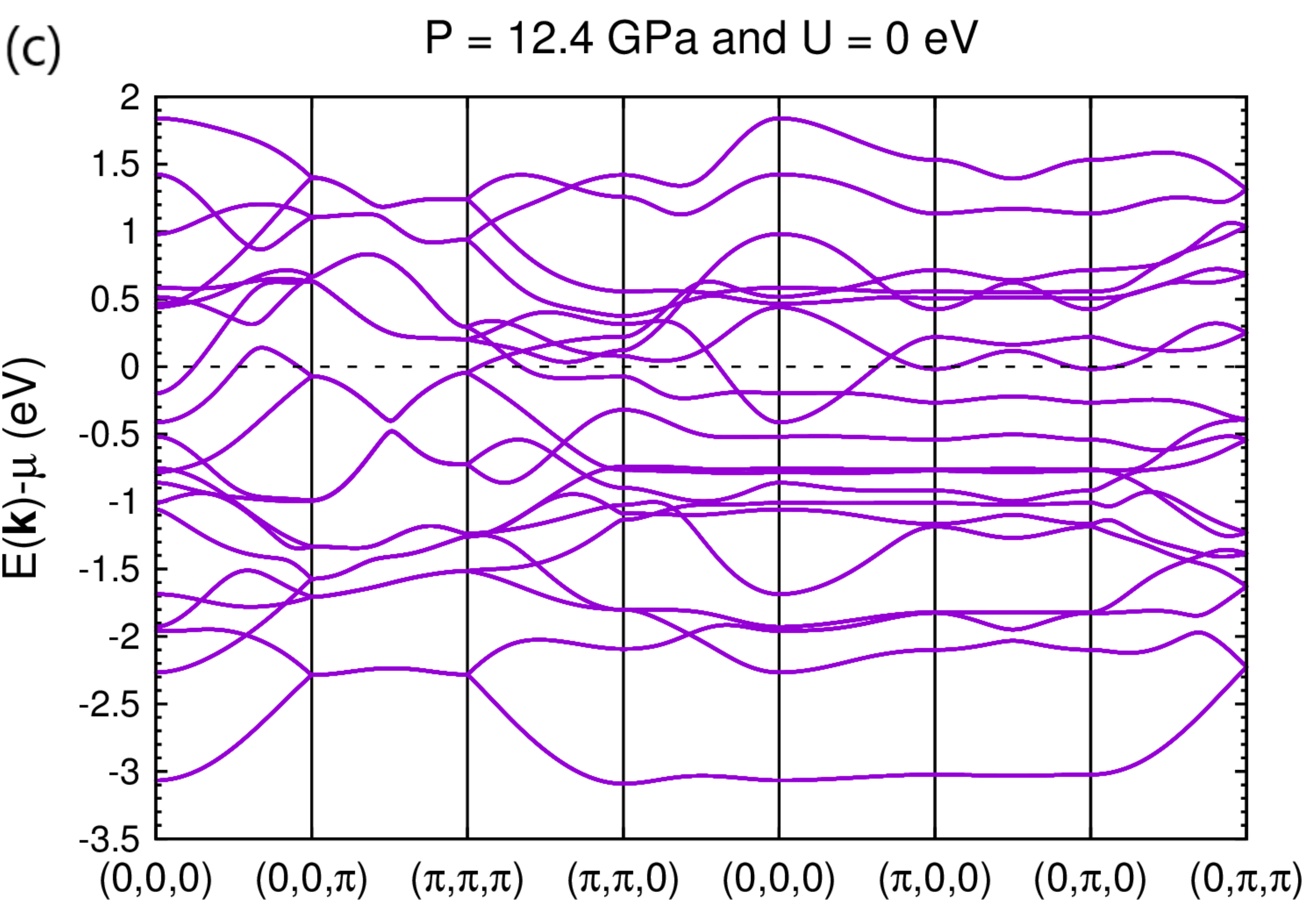} 
\includegraphics[clip,width=0.42\textwidth]{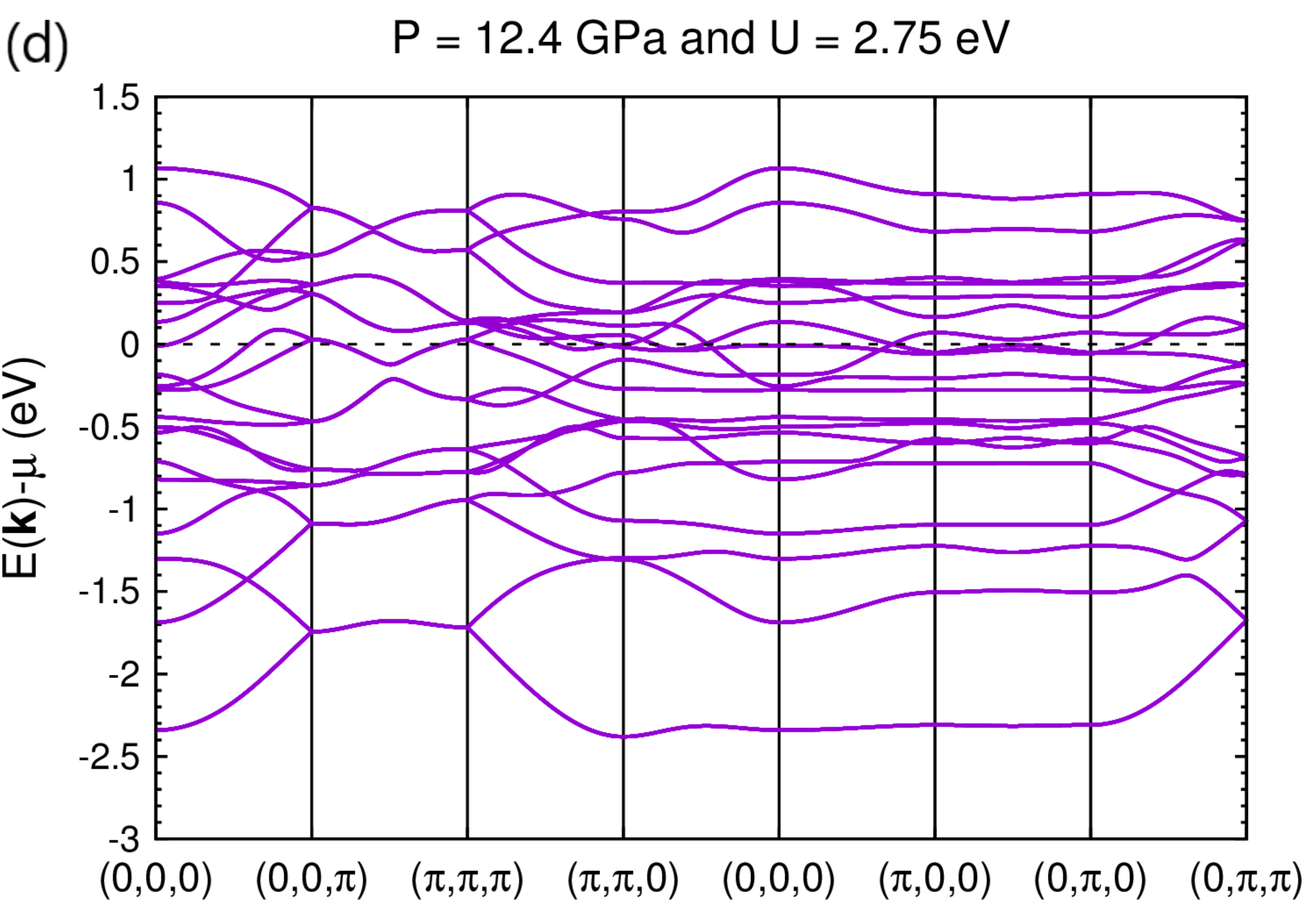} 
\caption{(Color online) Band structure for $P=0$ and $P=12.4$ 
GPa with and without including the electronic correlations.  The bandstructure in (a) and (c) for $P=0$ and $P=12.4$ GPa is calculated using the tight-binding model from\citep{arita_prb2015-downfolding} and mimics the one obtained in ab-initio calculations, see text. The bandwidth is approximately 4 eV and 5 eV respectively for $P=0$ and $P=12.4$ GPa. As shown in (c) and (d) the electronic correlations reduce these bandwidth below 2 eV and 3.5 eV, respectively. Band narrowing, is more evident close to the Fermi level } 
\label{fig:Figbandas} 
\vspace{-0.3cm}
\end{figure*}

\begin{figure*}
\leavevmode
%\hskip -1.0cm
\includegraphics[clip,width=0.42\textwidth]{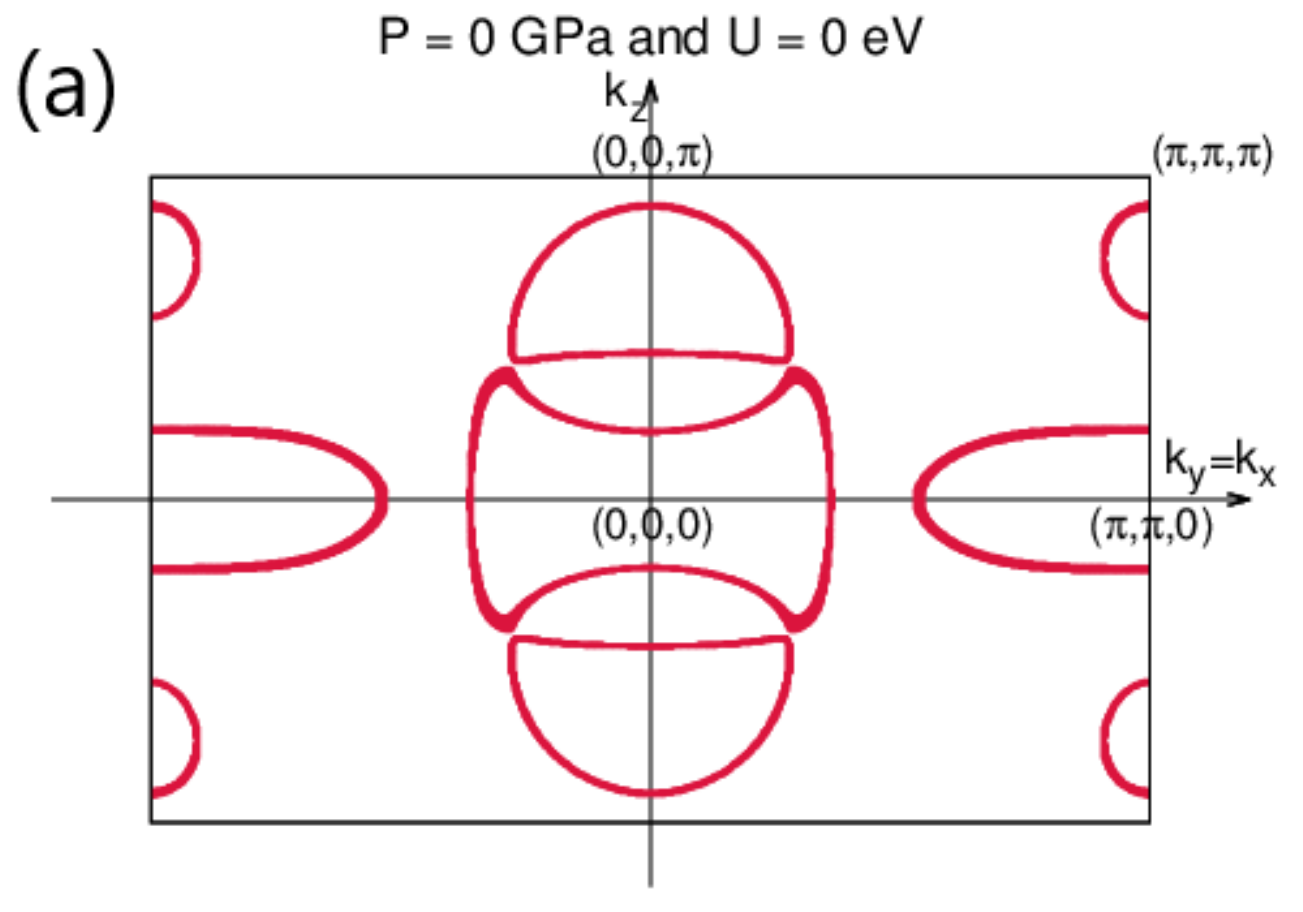} 
\includegraphics[clip,width=0.42\textwidth]{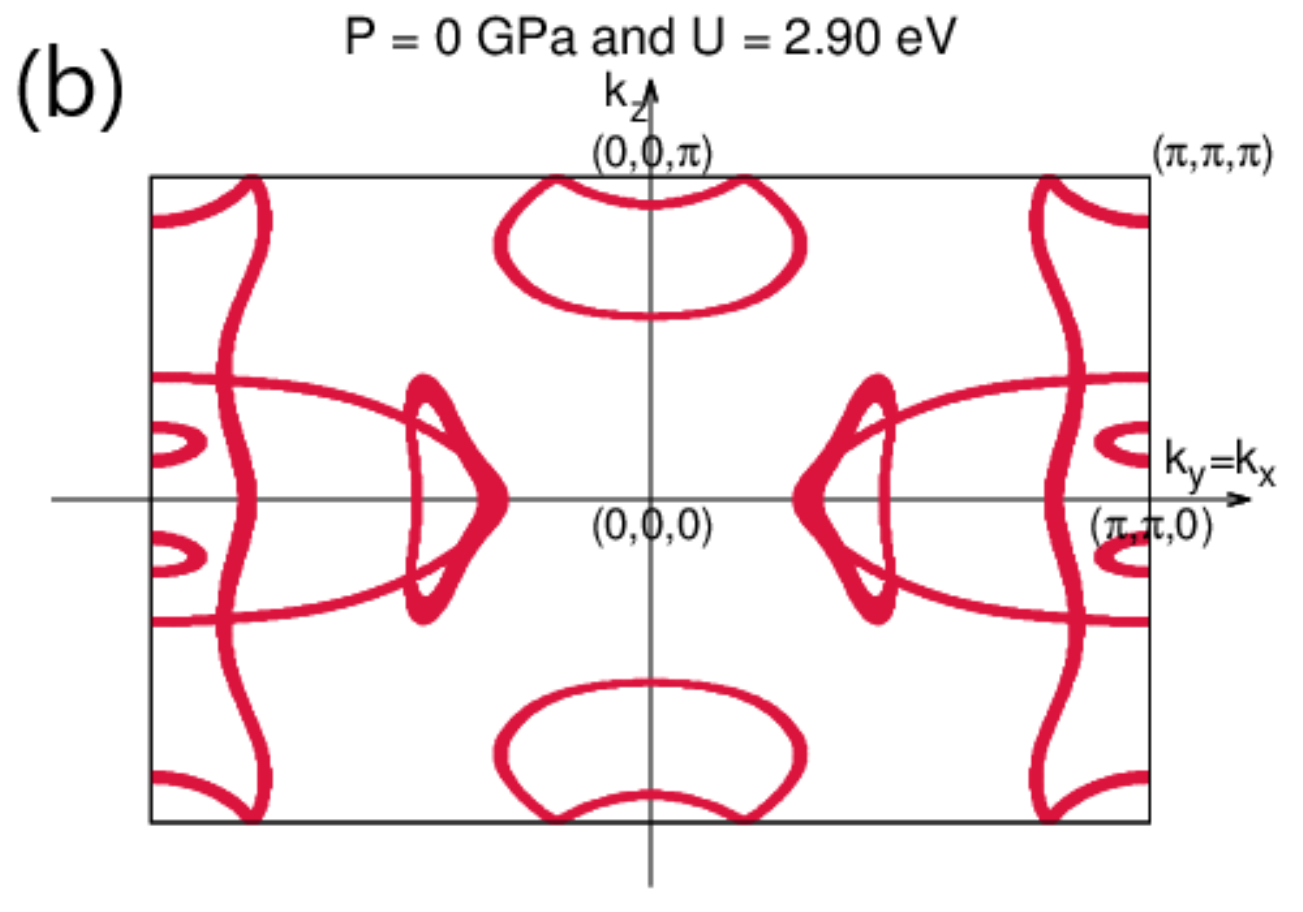} 
\includegraphics[clip,width=0.42\textwidth]{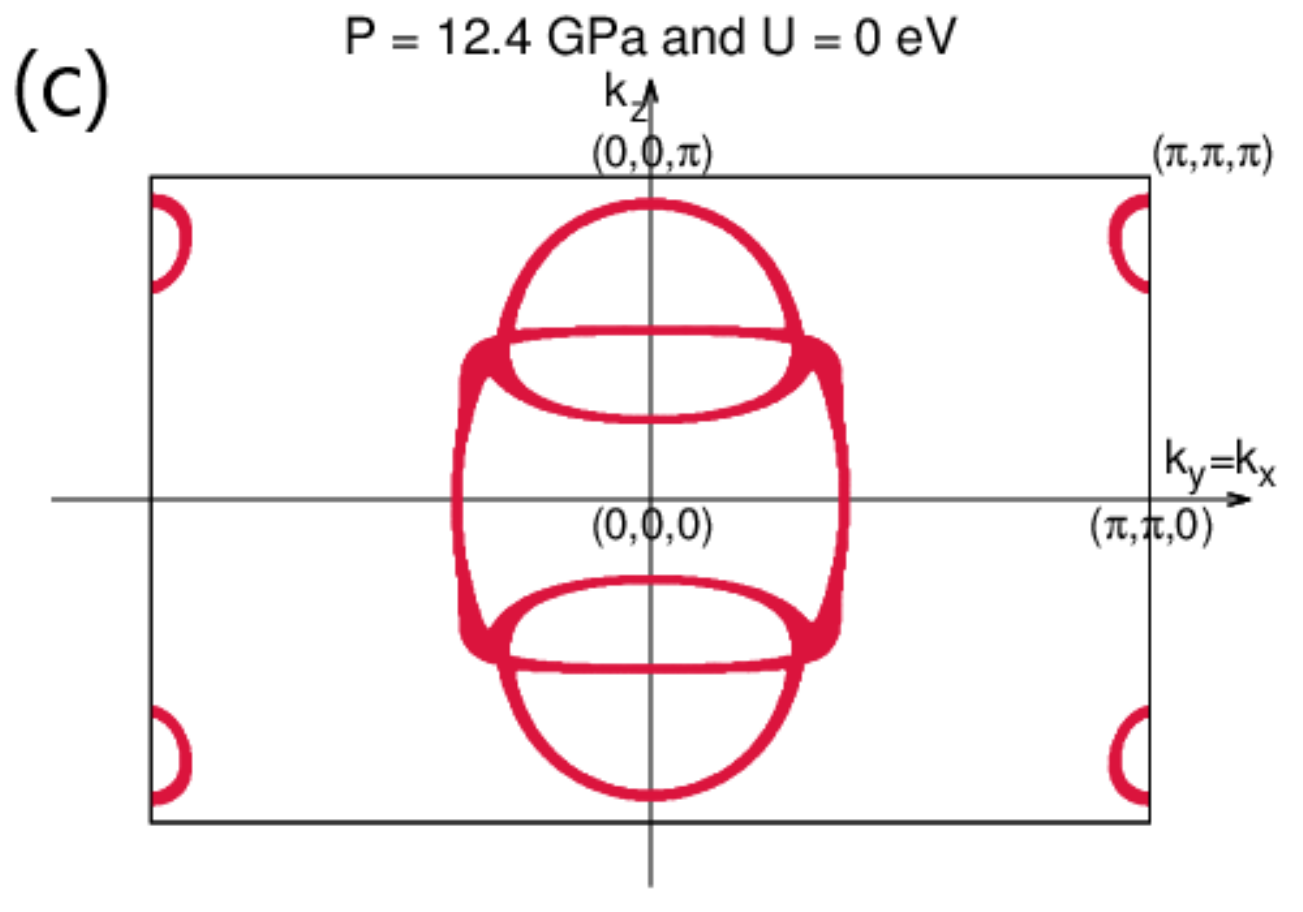} 
\includegraphics[clip,width=0.42\textwidth]{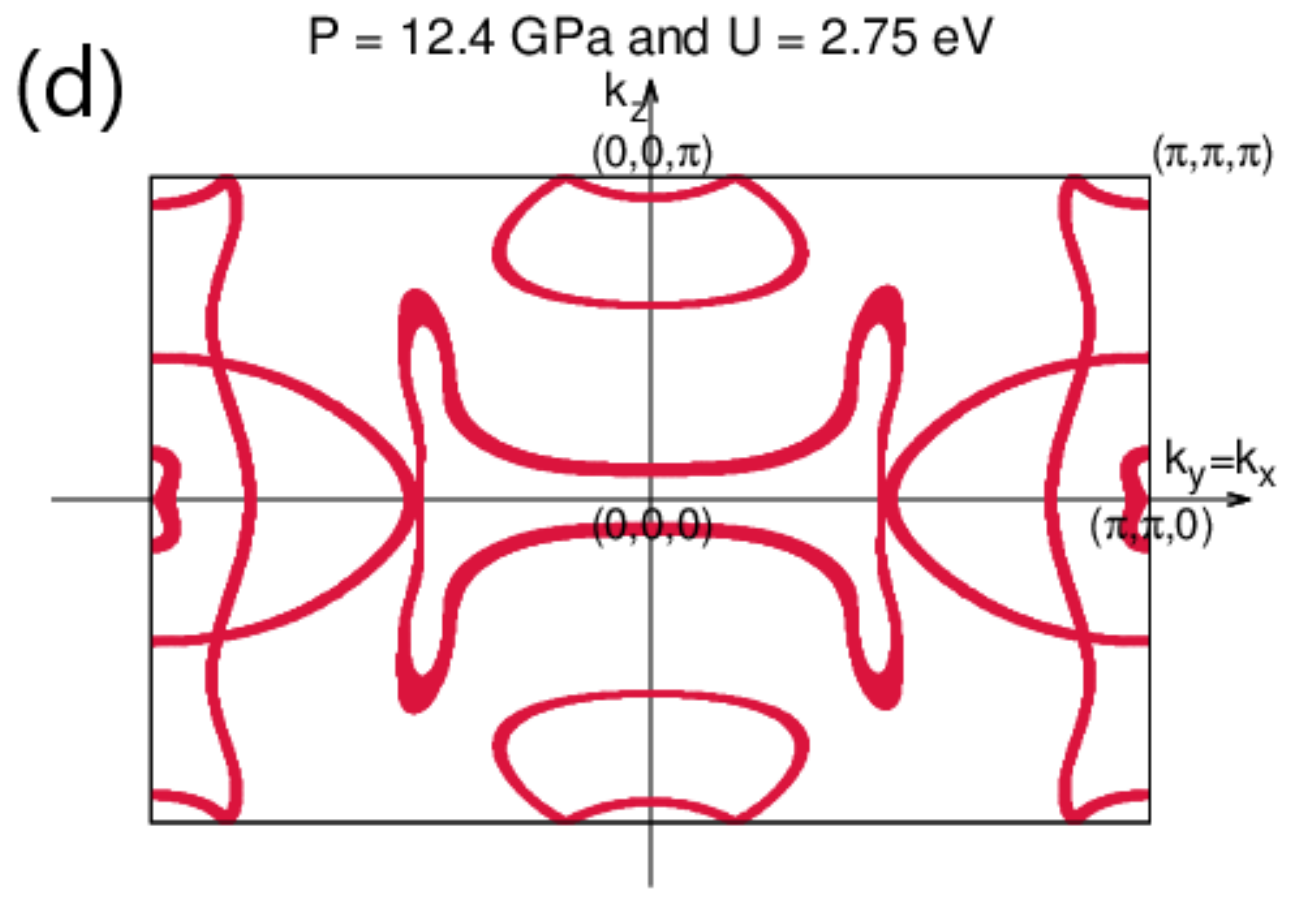} 
\caption{(Color online) Fermi surface cuts  for $P=0$ and $P=12.4$ 
GPa with and without including the electronic correlations  plane, along $k_z$-$k_x=k_y$ reciprocal to the Fe-ladder plane. At both pressures the Fermi surface is reconstructed due to electronic correlations.} 
\label{fig:Figcuts1} 
\vspace{-0.3cm}
\end{figure*}

\begin{figure*}
\leavevmode
%\hskip -1.0cm
\includegraphics[clip,width=0.42\textwidth]{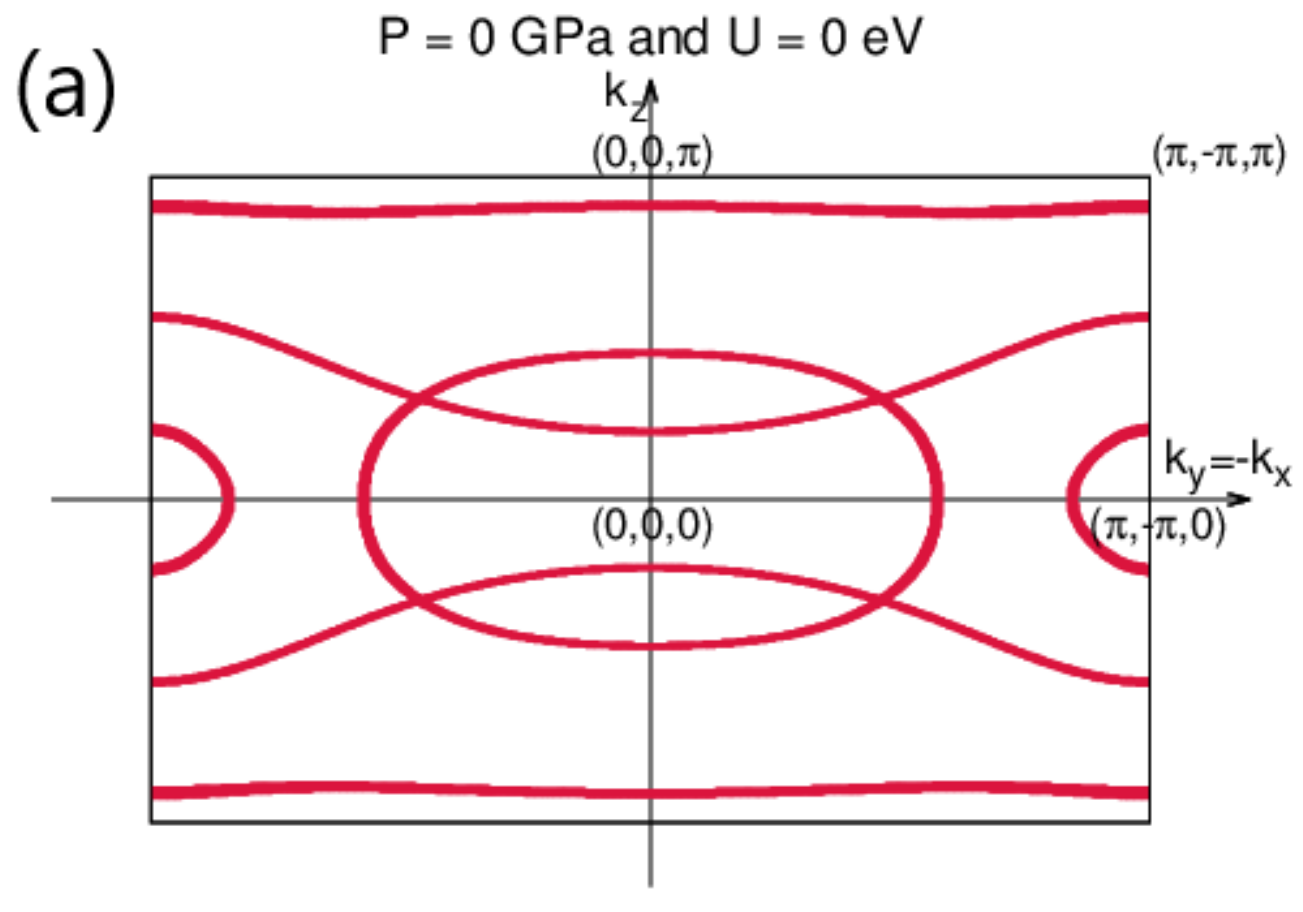} 
\includegraphics[clip,width=0.42\textwidth]{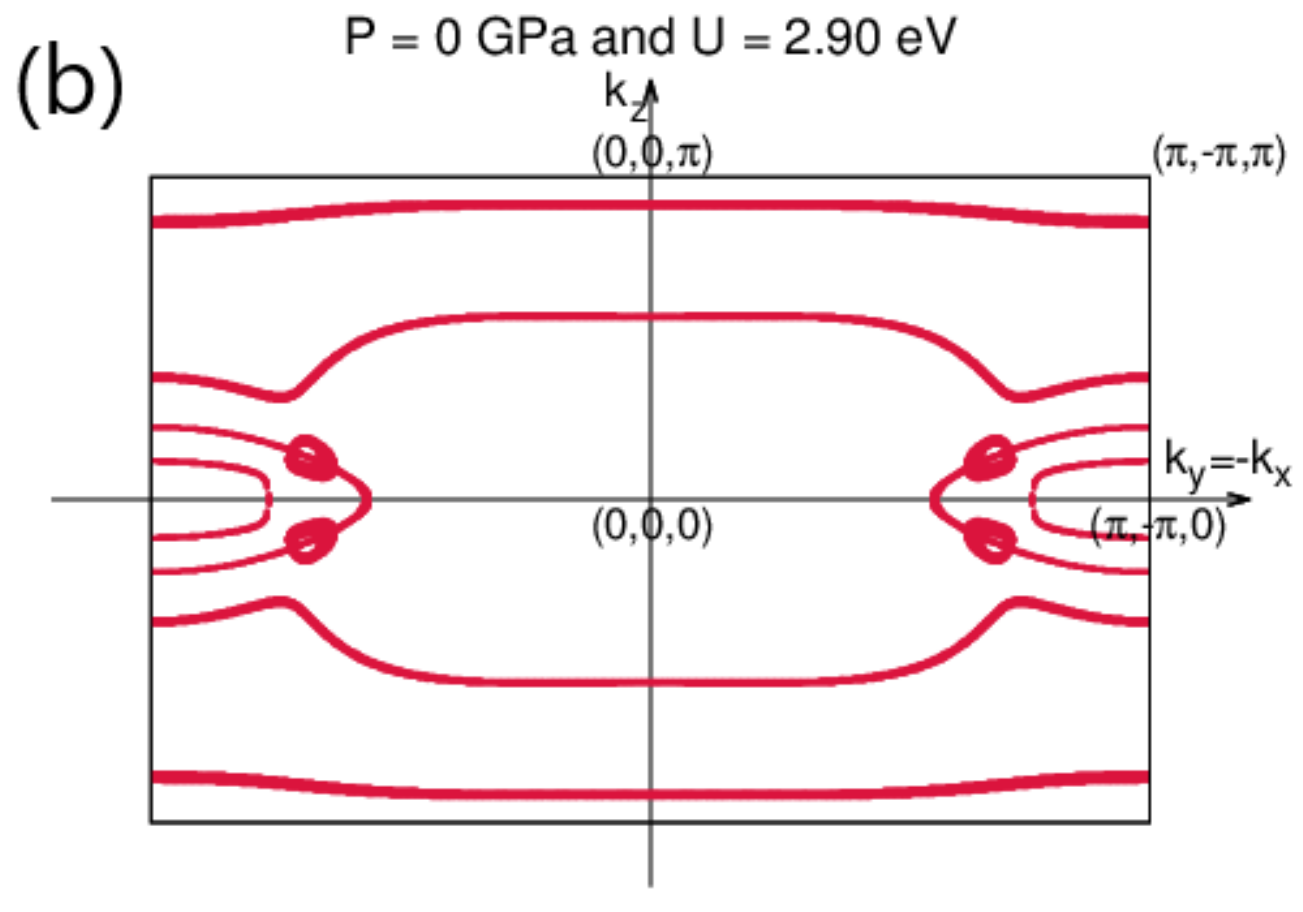} 
\includegraphics[clip,width=0.42\textwidth]{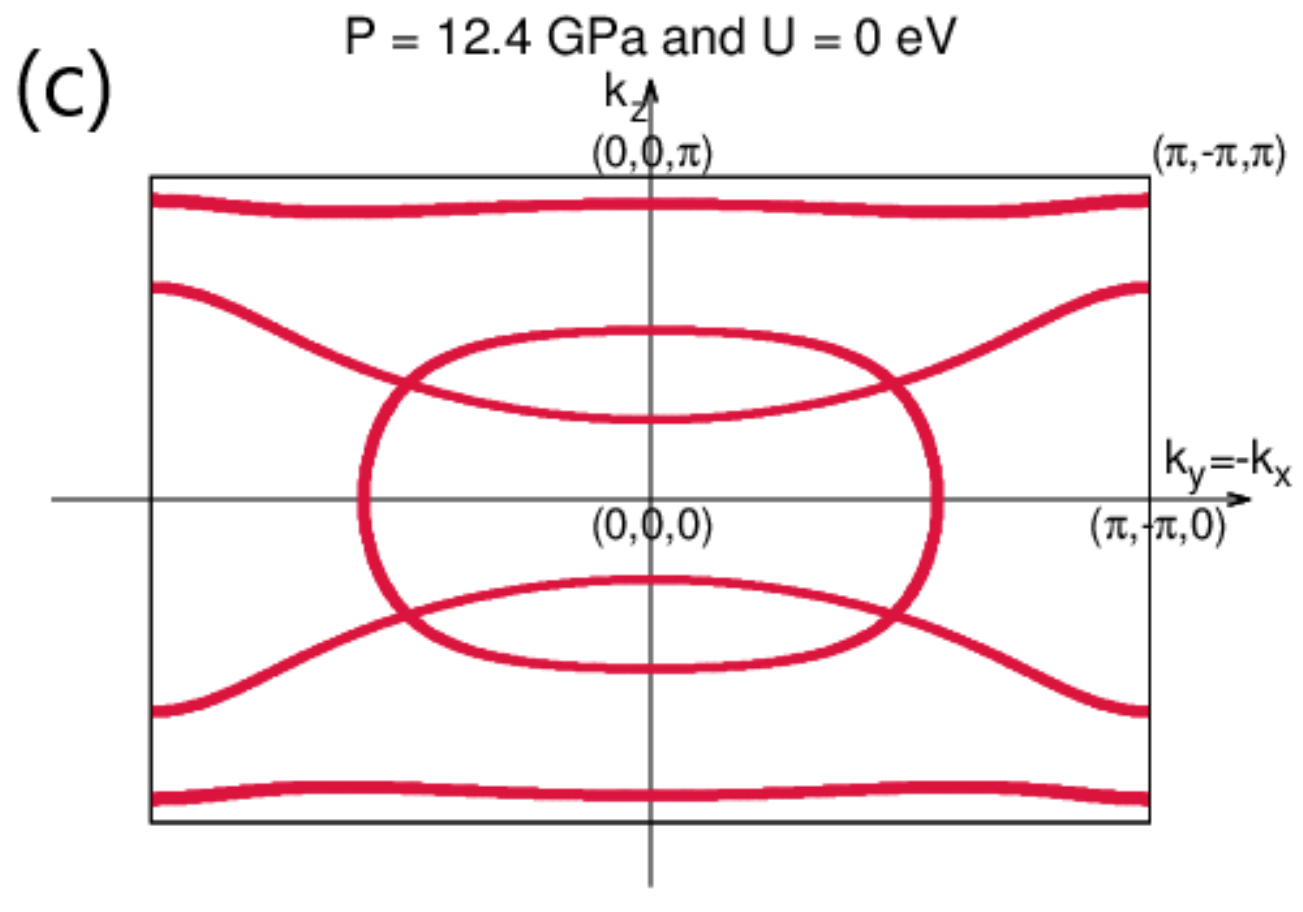} 
\includegraphics[clip,width=0.42\textwidth]{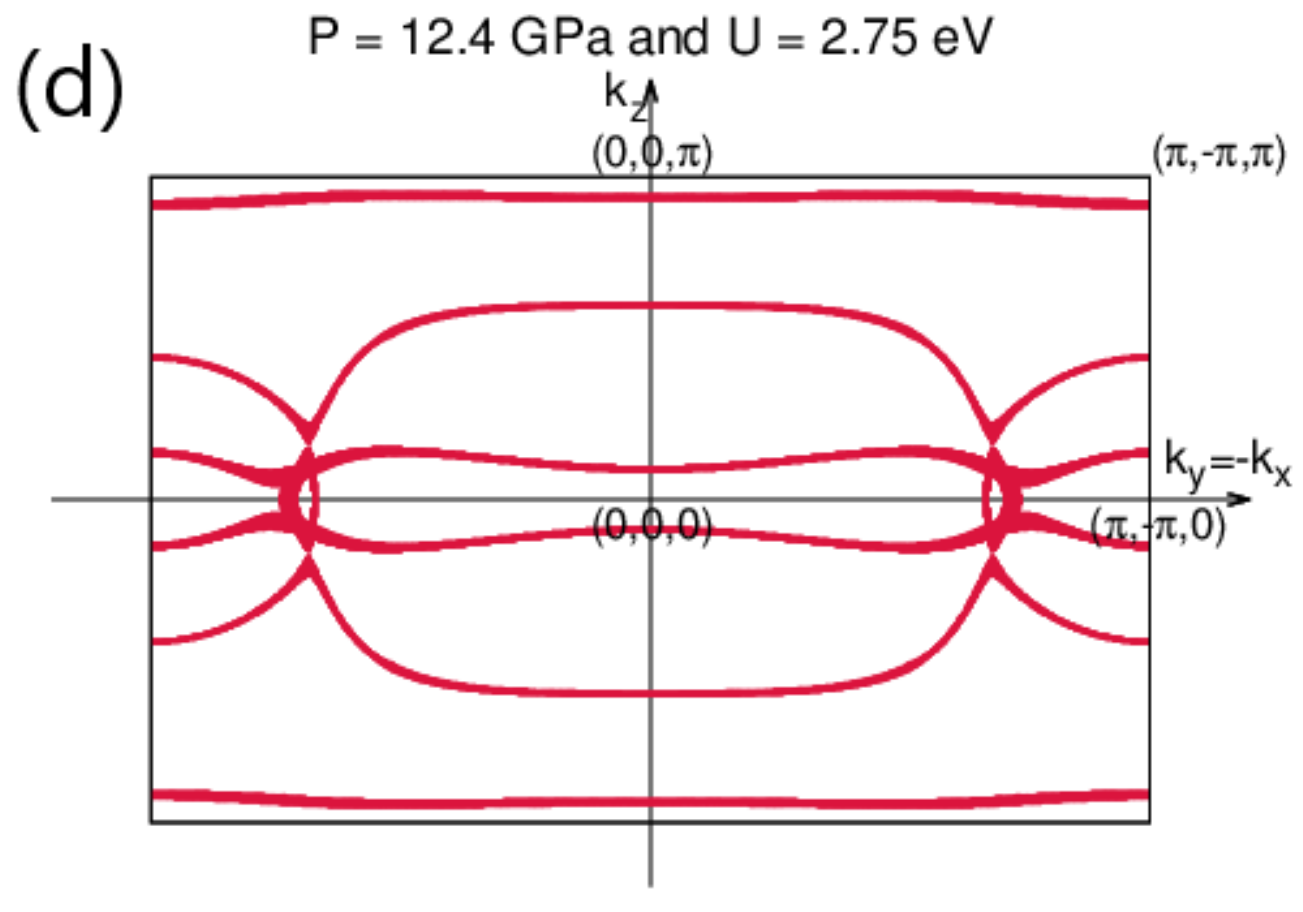} 
\caption{(Color online)Fermi surface cuts for $P=0$ and $P=12.4$ 
GPa with and without including the electronic correlations along $k_z$-$k_x=k_y$ plane, reciprocal to the plane perpendicular to the Fe-ladder plane } 
\label{fig:Figcuts2} 
\vspace{-0.3cm}
\end{figure*}

\end{document}